\documentclass[preprint,amsmath,amssymb,aps]{revtex4-1}
\usepackage{multirow}
\usepackage{graphicx}
\usepackage{dcolumn}
\usepackage{bm}
\begin{document}

\title{Sub-grid-scale model for studying Hall effects on macroscopic aspects of magnetohydrodynamic turbulence}  

\author{Hideaki Miura}
 \email{miura.hideaki@nifs.ac.jp}
\affiliation{%
 National Institute for Fusion Science, Toki, Gifu 509-5292, JAPAN
}%

\author{Fujihiro Hamba}%
\affiliation{
  Institute of Industrial Science, The University of Tokyo, 4-6-1 Komaba, Meguro-ku, Tokyo 153-8505, JAPAN
}%

\date{\today}

\begin{abstract}
A new sub-grid-scale model is developed for studying influences of the Hall term on macroscopic aspects of magnetohydrodynamic turbulence.
Although the Hall term makes numerical simulations extremely expensive by exciting high-wave-number coefficients and makes magnetohydrodynamic equations stiff, studying macroscopic aspects of magnetohydrodynamic turbulence together with the Hall term is meaningful since this term often influences not only sub-ion-scales but also macroscopic scales.
A new sub-ion-scale sub-grid-scale model for large eddy simulations of Hall magnetohydrodynamic turbulence is developed in order to overcome the difficulties.
Large eddy simulations by the use of the new model successfully reproduce statistical natures such as the energies and probability density functions of the vorticity and current density, keeping some natures intrinsic to Hall magnetohydrodynamic turbulence.
Our new sub-grid-scale model enables numerical simulations of homogeneous and isotropic Hall magnetohydrodynamic turbulence with a small computational cost, improving the essential resolution of an LES from that carried out with earlier models, and retaining the ion-electron separation effects by the Hall term in the grid scale.
\end{abstract}


\maketitle

\section{Introduction}
Magnetohydrodynamic (MHD) turbulence has been studied extensively for the purpose of understanding macroscopic plasma turbulence as well as understanding universal physics of turbulence which should be shared with neutral fluid turbulence.
Numerous theoretical, numerical, and observational works have been devoted to understand MHD turbulence over the past decades\cite{Iroshnikov1964,Kraichnan1965,SridharGoldreich1994,GoldreichSridhar1997,Stawicki2001,ChoLazarian2002,Bale2005,Biskamp2003,Horbury2008,Alexandrova2009,Kiyani2009,Servidio2009,Davidson2013,Bruno2016}.
More recently, a study of sub-ion-scale effects on turbulence has gathered wide attention\cite{Mahajan2000,Matthaeus2003,Smith2004,Ren2005,Dmitruk2006,Galtier2007,Mininni2007,Galtier2008,HoriMiura2008,Miura2013,Miura2014a}.
The Hall term represents effects of the ion-electron separation in the electric field, the scale of which can be characterized by the ratio of the ion skin-depth to the system size (Hall parameter).
The Hall term can change turbulence dynamics at the small, near-dissipation ion-skin-depth scale, especially through its modified frozen-in condition of the magnetic field to the velocity field.

The Hall effects have been discussed as modifying the scaling-law of turbulence to $k^{-7/3}$ or $k^{-11/3}$ ($k$ is the wave number)\cite{Galtier2008,HoriMiura2008,Galtier2012,Miura2013} and coherent turbulent structures\cite{Dmitruk2006,Miura2014a,Miura2019}.
The Hall term appears also in studies of basic plasma instabilities\cite{Huba1996,Winske1996,Zhu2008,Gotor2014,Gotor2015,Umeda2017} as well as of nuclear fusion\cite{Park1999,Sugiyama2000,Sovinec2004,Miura2014b} which are studied by the use of extended MHD equations\cite{Braginskii1965,HazeltineMeiss1991,Schnack2006}.

Since plasma dynamics in a small scale can be closely coupled with larger scales through advection and other nonlinear terms, introducing the Hall or two-fluid effects at the small ion skin-depth scales can influence dynamics in larger scales.
In fact, our earlier studies in Ref.\cite{Miura2014a,Miura2019} have revealed that the Hall term can induce a structure transition of vortices involving scales larger than the ion skin depth, making the turbulent field be thoroughly different from that of non-Hall MHD (or single-fluid MHD) turbulence qualitatively.
Furthermore, the Hall term can affect large-scale magnetic field dynamics through influencing local magnetic reconnection events in turbulent field\cite{Servidio2009} or through other kinetic effects\cite{Umeda2017}.

The Hall term not only enriches physics such as whistler waves but also increases a numerical difficulty in simulations.
A time step width $\Delta t$ in a simulation is limited strictly as $\Delta t < 1 /  k_{\parallel} k$ ($k$ and $k_{\parallel}$ are the wave number and wave number parallel to magnetic field, respectively) in a Hall MHD simulation because of the whistler waves.
In addition, the Hall term can enhance some instabilities at small scales, obliging us using either a large number of grid points or a hyper-diffusivity in numerical simulations\cite{Huba1996,Winske1996,Zhu2008,Gotor2014,Gotor2015}.
Thus Direct Numerical Simulations (DNS) of Hall MHD turbulence are much more expensive than DNS of non-Hall MHD equations. 
Furthermore, with respect to a real application such as numerical simulations of instabilities in a magnetically confined nuclear fusion, an excessively large computational cost is not very fruitful because large-scale motions are of main interest.
Nevertheless, since Hall effects can influence a scale larger than the ion skin depth, we must take the Hall effects into large-scale dynamics while suppressing the computational cost.

For studies of large-scale turbulence, so-called Large Eddy Simulations (LES) can be carried out at a much smaller computational cost than DNS\cite{Durbin2001,Garnier2009}.
In LES, we operate a low-pass filter to a set of governing equations, and replace effects of the scales smaller than the filter width by a phenomenological model while retaining the scales larger than the filter width (or the Grid Scale, GS).
Such a phenomenological model is called the Sub-Grid-Scale(SGS) model.
Spatial and temporal resolutions of LES of Hall MHD turbulence can be much coarser than DNS, with an expense of precise dynamics near to or smaller than the filter width.

While LES of MHD turbulence have been studied in some earlier works\cite{Kobayashi2006,Petrosyan2006,Hamba2010,David2012} (see also Refs.\cite{Petrosyan2014,Miesch2015} and references therein), studies of LES of Hall MHD turbulence are not very popular.
Recently, we have shown in Ref.\cite{Miura2016} that the LES approach can be applicable to homogeneous and isotropic Hall MHD turbulence as well as homogeneous turbulence under a constant and uniform magnetic field by applying the SGS model developed in Ref.\cite{Hamba2010}.
We have also applied the LES approach to numerical simulations of nonlinear evolution of ballooning instability in a heliotron-type device for magnetically confined nuclear fusion experiments\cite{Miura2017} by adopting the same type of SGS model as in Ref.\cite{Miura2016} to Braginskii-type extended MHD equations.
More recently, Camporeale et al.\cite{Camporeale2018} analyzed coherent structures in 2D Hall MHD turbulence simulations by the use of the filtering technique, mentioning the LES approach explicitly. 
These numerical works have shown the effectiveness of the LES approach of Hall MHD/extended MHD simulations.

In spite of the success in Ref.\cite{Miura2016,Miura2017}, there remains a subject to be studied further on SGS modeling of the Hall term.
We have taken the Smagorinsky constants of the SGS models relatively large, with an expense of a reduced numerical resolution (a steep decay of the energy spectra), in order to reproduce the coherent or local structures of turbulence such as vortex tubes and current sheets clearly.
Since reproduction of both the energy spectra and the local structures are important, we need to find a better compromise between them.
The purpose of this paper is to develop a new SGS model for Hall MHD equations in order to improve the compromise and study natures of the model for homogeneous and isotropic turbulence.

This paper is organized as follows.
In \S.2, Hall MHD equations, the grid-scale (GS) equations of the Hall MHD model, and SGS model for Hall MHD turbulence are introduced.
In \S.3, LES of homogeneous and isotropic Hall MHD turbulence are carried out.
Results of the LES are compared to DNS results.
\S.4 is for summary.

\section{Homogeneous Hall MHD turbulence}
The incompressible Hall MHD equations can be expressed as
\begin{eqnarray}
  {\partial u_i \over \partial t} 
     & = &  - {\partial \over \partial x_j}\left[
                \left( u_i u_j  -B_i B_j\right)
            +  \left(p + {1 \over 2} B_k B_k \right) \delta_{ij}\right]
            + \nu {\partial S_{ij} \over \partial x_j},
    \label{eq:HallMHD1}
     \\
 {\partial u_k \over \partial x_k } & = &  0, 
    \label{eq:HallMHD2}
     \\
 S_{ij} & =& {\partial u_i \over \partial x_j}
           + {\partial u_j \over \partial x_i}, 
    \label{eq:HallMHD3}
    \\
 {\partial B_i \over \partial t}
     & = &  - \epsilon_{ijk} {\partial E_k \over \partial x_j}, 
    \label{eq:HallMHD4}
    \\
 {\partial B_k \over \partial x_k } & = &  0,
    \label{eq:HallMHD5}
    \\
  E_i & =& -\epsilon_{ijk} 
              \left( u_j - \epsilon_H J_j \right) B_k + \eta J_i. 
    \label{eq:HallMHD6}
\end{eqnarray}
The three symbols $B_i$, $J_i = \epsilon_{ijk} \partial_j B_k$, and $u_i$ represent the $i$-th components of the magnetic field, the current density, and velocity field vectors, respectively.
Tensor symbols $\delta_{ij}$ and $\epsilon_{ijk}$ are the Kronecker's delta and the Levi-Civita's anti-symmetric tensor, respectively.
The sum of $1$, $2$, and $3$ is taken for repeated suffixes of the vector and tensor variables.
The equations (\ref{eq:HallMHD1})-(\ref{eq:HallMHD6}) are already normalized by a reference length scale $L_0$, magnetic field strength $B_0$, mass density $\rho_0 = m_0 n_0$ ($m_0$ and $n_0$ are the ion mass and the number density, respectively), and Alfv{\'e}n velocity $V_A=\sqrt{B^2/\mu_0 \rho_0}$ where $\mu_0$ is the vacuum permeability.
The symbols $\epsilon_H$, $\eta$, and $\nu$ are the Hall parameter (the ratio of the ion skin depth $d_i = \sqrt{m_0/n_0 \mu_0 e^2}$ to the length scale $L_0$, where $e$ is the electric charge), magnetic diffusivity, and shear viscosity, respectively.
We can understand $1/\nu$ and $1/\eta$ as the reference Reynolds number and the Lundquist number, respectively. 
These parameters control near-dissipation scales of turbulence in simulations.

The equations (\ref{eq:HallMHD1})-(\ref{eq:HallMHD6}) are operated by a low-pass filter
\begin{eqnarray}
  {\overline {v}} & := & \int G(x-s) v(s) ds
\end{eqnarray}
to obtain the GS equations of the Hall MHD model, where $v(x,t)$ is a dependent variable.
The filter $G$ is assumed to be a sharp and isotropic in the Fourier space, with the cut-off wave number $k_{cut}$.
The GS Hall MHD equations are described as
\begin{eqnarray}
  {\partial {\overline u}_i \over \partial t} 
     & = &  - {\partial \over \partial x_j}\left[
           \left( {\overline u}_i {\overline u}_j 
                 -{\overline B}_i {\overline B}_j\right)
            +  \left ({\overline p} 
             + {1 \over 2}{\overline B}_k {\overline B}_k \right) \delta_{ij}
            \right]
            + \nu {\partial {\overline S}_{ij} \over \partial x_j} 
          - {\partial {\overline \tau}_{ij} \over \partial x_j},
    \label{eq:GSHallMHD1} \\
 {\overline \tau}_{ij} 
 & = &  \left[
    \left( {\overline {u_i u_j} } -{\overline {B_i B_j}}\right)
        + {1 \over 2} {\overline {B_k B_k}} \delta_{ij} \right]
 -  \left[
    \left( {\overline u}_i {\overline u}_j -{\overline B}_i {\overline B}_j\right)
        +  {1 \over 2}{\overline B}_k {\overline B}_k \delta_{ij} \right],
    \label{eq:GSHallMHD2}  \\
 {\overline S}_{ij} & =& {\partial {\overline u}_i \over \partial x_j}
           + {\partial {\overline u}_j \over \partial x_i}, 
    \label{eq:GSHallMHD4}
    \\
 {\partial {\overline u}_k \over \partial x_k } & = &  0, 
    \label{eq:GSHallMHD3}  \\
 {\partial {\overline B}_i \over \partial t}
    & = &  - \epsilon_{ijk} {\partial \overline{E}_k \over \partial x_j}, 
    \label{eq:GSHallMHD5} \\
{\overline {E}}_i 
    & =& -\epsilon_{ijk} \left( \overline {u}_j 
           - \epsilon_H {\overline J}_j \right) \overline {B}_k 
           + \eta \overline{J}_i 
         - {\overline E}^M_i - {\overline E}^H_i, 
    \label{eq:GSHallMHD6} \\
\overline {E}^M_i 
    & = & -\epsilon_{ijk} \left( 
            \overline{ u_j B_k } - \overline {u}_j \overline
                     {B}_k\right), 
    \label{eq:GSHallMHD7} \\
\overline {E}^H_i 
    & = & - \epsilon_H \epsilon_{ijk} \left( 
        -\overline{ J_j B_k } + \overline {J}_j \overline
        {B}_k\right),
     \label{eq:GSHallMHD8} \\
 {\partial \overline {B}_k \over \partial x_k} & = &  0.
     \label{eq:GSHallMHD9}
\end{eqnarray}

The GS equations (\ref{eq:GSHallMHD1})-(\ref{eq:GSHallMHD8}) are closed when the SGS terms ${\overline \tau}_{ij}$, $\overline{E}^M_i$, and $\overline{E}^H_i$ are replaced by SGS models expressed only by the GS variables $\overline{u}_i$ and $\overline{B}_i$.
In Ref.\cite{Miura2016}, the SGS terms $\overline{\tau}_{ij}$, $\overline {E}^M_i$ and $\overline {E}^H_i$ have been modelled as
\begin{eqnarray}
  \overline{\tau}_{ij} & = & - \nu_{{}_{SGS}} \overline{S}_{ij},                \label{eq:SGS1}
\\
  \overline{E}^M_i +\overline{E}^H_i & = & - \eta_{{}_{SGS}} \overline{J}_i,    \label{eq:SGS2}
\\
  \nu_{{}_{SGS}}
      & =& C_{\nu}\Delta^2 \left({1 \over 2} C_{\nu} \overline{S}_{ij}^2 + C_{\eta} {\overline J}_i^2 \right)^{1/2}, \label{eq:SGS3}
\\ 
  \eta_{{}_{SGS}} & =& C_{\eta}\Delta^2 \left({1 \over 2} C_{\nu} \overline{S}_{ij}^2 + C_{\eta} {\overline J}_i^2 \right)^{1/2}, \label{eq:SGS4}
\end{eqnarray}
based on the studies in Refs.\cite{Hamba2010,Miura2013,Miura2014a}.
The symbol $\Delta$ is the filter width.
In this model, we have two Smagorinsky constants $C_{\nu}$ and $C_{\eta}$.
They have been given originally in Ref.\cite{Hamba2010} as $C_{\nu}=0.046$ and $C_{\eta}=C_{\nu}\times (5/7)$.
However, in Ref.\cite{Miura2016}, we have chosen coefficients $C_{\nu}=C_{\eta}=0.345$ for $\Delta = \Delta_0 = 1/k_{cut}$ to reproduce local structures, because the turbulent field is too noisy when the Smagorinsky constants are small.
Consequently, the numerical resolution in the LES has been reduced effectively.

In order to find a better compromise between the reproduction of the energy spectra and spatial structures, we model $\overline{E}^M_i$ and $\overline{E}^H_i$ as
\begin{eqnarray}
  \overline{E}^M_i + \overline{E}^H_i 
  & = & - \eta_{{}_{SGS}} \overline{J}_i \nonumber \\
  &   & + \epsilon_H {\partial \over \partial x_j}
     \left( \nu_{{}_{SGS}} \sigma_{{}_{H,1}} \overline{S}_{ij} \right)
     - \epsilon_H^2 {\partial \over \partial x_j}
     \left[ \nu_{{}_{SGS}} \sigma_{{}_{H,2}} \overline{S}_{ij}
       \left( {\partial {\overline J}_i \over \partial x_j}
            + {\partial {\overline J}_j \over \partial x_i}\right)
     \right],
  \label{eq:SGS8}
\end{eqnarray}
keeping $\overline{\tau}_{ij}$ the same as eq.(\ref{eq:SGS1}).
We do not separate $\overline{E}^M_i$ and $\overline{E}^H_i$ because a forward-transfer part of the Hall term in the Fourier space can be combined as the diffusive part of $\overline{E}^H_i$ together with the $\overline{E}^M_i$.
See Appendix for more detailed description on the model.

The contribution of this SGS model to the magnetic energy budget ${\partial_t}({1 \over 2} B_i B_i)$ is
\begin{eqnarray}
  -\epsilon_{MSH}
  & = & -\overline{E}^H_i \overline{J}_i \nonumber \\
  & = & \epsilon_H 
     \left( \nu_{{}_{SGS}} \sigma_{{}_{H,1}} \overline{S}_{ij} \right)
     \left( {\partial {\overline J}_i \over \partial x_j} \right)
     - \epsilon_H^2
       \nu_{{}_{SGS}} \sigma_{{}_{H,2}} 
       \left( {\partial {\overline J}_i \over \partial x_j}
            + {\partial {\overline J}_j \over \partial x_i}\right)
       \left( {\partial {\overline J}_i \over \partial x_j} \right) \nonumber \\
   && - {\partial \over \partial x_j}\left[
     \epsilon_H 
     \left( \nu_{{}_{SGS}} \sigma_{{}_{H,2}} \overline{S}_{ij} \right)
     {\overline J}_i 
     - \epsilon_H^2 
     \nu_{{}_{SGS}} \sigma_{{}_{H,2}}
     \left( {\partial {\overline J}_i \over \partial x_j}
            + {\partial {\overline J}_j \over \partial x_i}\right)
       \overline{J}_i \right].
\end{eqnarray}
While the second term works only for the energy dissipation and the third term vanishes by the volume integral over the computational domain, a behavior of the first term depends on detailed dynamics in the nonlinear evolution.

%
%

\section{LES of decaying homogeneous and isotropic Hall MHD turbulence}
\subsection{Outline of LES}
LES of freely decaying homogeneous and isotropic Hall MHD turbulence for ${\epsilon}_H = 0.05$ and ${\epsilon}_H=0.025$ are carried out in this section.
The LES results are compared to the GS components of DNS data reported in Ref.\cite{Miura2014a}, with the same viscosity, resistivity, and Hall parameter, and the same initial condition as the LES.
The initial velocity and magnetic fields are given by the energy spectrum $E(k,t) \propto k^2 \exp(-k^2/k_0^2)$ ($k_0=2$) and random phases of the Fourier components.
Equations~(\ref{eq:GSHallMHD1})-(\ref{eq:GSHallMHD8}) are coupled with with the SGS model in eqs.~(\ref{eq:SGS1}), (\ref{eq:SGS3}), (\ref{eq:SGS4}), and (\ref{eq:SGS8}), and solved numerically by the pseudo-spectral method and the Runge-Kutta-Gill scheme.
Aliasing errors are removed by the $2/3$-truncation in spectral space.
Note that the implementation of the $2/3$-truncation is different between this paper and Ref.\cite{Miura2016}.
The $2/3$-truncation has been implemented in Ref.\cite{Miura2016} as anisotropic low-pass filter (rectangular in the spectral space), in order to study not only homogeneous and isotropic turbulence but also homogeneous turbulence under a constant and uniform magnetic field.
In this paper, to the contrary, an isotropic filter (spherical in the spectral space) is adopted for de-aliasing because we focus on homogeneous and isotropic turbulence.

An advantage of a Hall MHD simulation to a (non-Hall) MHD simulation with respect to LES is that the forward energy transfer of the magnetic energy can be dominant at a large wave number region in the spectral space, because the Hall term makes the magnetic field (induction) equations be quadratic to the magnetic field\cite{Miura2013,Miura2014a}.
In order to make use of this advantage, we study firstly LES of $k_{max} > k_H = 1/\epsilon_H$ so that the GS Hall MHD equations retain the scales comparable to or smaller than the ion skin depth $\epsilon_H$, where $k_{max}$ is the maximum wave number in a simulation.
We will also examine LES with $k_{max} \simeq k_H$ below.
We refer to the DNS data with the number of grid points $N^3=1024^3$ and the parameters $\eta = \nu = 5 \times 10^{-4}$, $\epsilon_H = 0.05$, and $0.025$ reported in \cite{Miura2014a} as reference data.
Since we adopt the isotropic low-pass filter for de-aliasing, the maximum wave number in an LES is $k_{max} = 42$ for $N^3=128^3$ and $k_{max} = 21$ for $N^3=128^3$, while the wave number associated with the ion skin depth is $k_H = 20$ for $\epsilon_H=0.05$ and $k_H = 40$ for $\epsilon_H=0.025$.

All the parameters needed for the SGS model studied in this paper are shown in Table \ref{tab:params}.
We place a higher priority on verifying applicability of our new SGS model to Hall MHD turbulence, and on clarifying the nature of the model, rather than on calibrating model constants included in the SGS model to reproduce some typical quantities precisely.
While we have changed $C_{\nu}$ and $C_{\eta}$ by keeping $\Delta=\Delta_0$, we can understand that the $C_{\nu}$ and $C_{\eta}$ values are given for a larger $\Delta$ (this means that the GS is coarser than that of $\Delta=\Delta_0$).
Thus we show $C_{\nu}$ and $C_{\eta}$ for $\Delta_0$,  $1.5 \Delta_0$, $2 \Delta_0$, and $3 \Delta_0$. 
Different combinations of $\left(C_{\nu}, C_{\eta}, \Delta\right)$ in each of parameter sets (000, 001, ...) in Table \ref{tab:params} are equivalent to each other and give the same results for the SGS model in eqs.(\ref{eq:SGS1})-(\ref{eq:SGS4}).

\subsection{LES with the previous SGS model}
Firstly, we carry out LES with $N^3=128^3$ and $\epsilon_H=0.05$, and thus $k_{max} < k_H$, for parameter sets 000-004, in order to clarify the subject to be studied in this paper.
We emphasize here again that the parameter sets 000-004 are the same as those studied in our earlier work\cite{Miura2016} but the de-aliasing filter is different in the LES.
Both $\sigma_{{}_{H,1}}$ and $\sigma_{{}_{H,2}}$ are null in the parameter sets 000-004, and thus the Hall SGS term which is developed newly in this paper is switched off in the runs with the parameter sets 000-004 (hereafter referred to as runs 000-004).

In Fig.\ref{fig:LESstats01}, (a) the mean kinetic energy $E_K = \left< u_i u_i\right>/2 $, (b) the mean magnetic energy $E_M = \left< B_i B_i\right>/2$, (c) the enstrophy $Q = \left< \omega_i \omega_i\right>/2$ where $\omega_i = \epsilon_{ijk} \partial_j u_k$ is the $i$-th component of the vorticity, and (d) the mean current $J = \left < J_i J_i \right> /2$ are shown for the LES with the parameter set 000-004, together with the GS component of $E_K$ obtained from the reference DNS data filtered by $k_{cut}=42$ low-pass filter.
Hereafter, we refer to DNS data filtered as DNS GS data.
The filter width is either $k_{cut}=42$ or $21$ low-pass filter corresponding to $k_{max}$ of LES that are to be compared with the DNS GS data.
The symbol $\left< \cdot \right>$ indicates the volume average.
In Fig.\ref{fig:LESstats01}(a), the parameter set 000 gives the best fitting of $E_K$ to that of the DNS GS data (thick line) among the parameter sets 000-004.
Run 002 overestimates $E_K$ while the other runs underestimate the time evolution of $E_K$.
In Fig.\ref{fig:LESstats01}(b), the parameter set 002 gives the best fit to the time evolution of $E_M$ in the GS, while 000 and the other parameter sets underestimate $E_M$.
The enstrophy $Q$ and mean current $J$ in the GS are in between those given by the parameter sets 000 and 003 in Fig.\ref{fig:LESstats01}(c) and (d), respectively.
The quantities $Q$ and $J$ in the runs 000 and 002 can be twice as large as the DNS GS data.

In Fig.\ref{fig:LESspec01}, (a) the kinetic energy spectra $E_K(k,t) = \sum_{\left[k\right]} {\widetilde u}_i(k,t) \widetilde {u}_i^{*}(k,t)$ and (b) the magnetic energy spectra $E_M(k,t) = \sum_{\left[k\right]} {\widetilde B}_i(k,t) {\widetilde B}_i^{*}(k,t)$ at $t=1$ are shown, where the symbols  $\sum_{\left[k\right]}$, ${\widetilde {\cdot}}$, and ${}^{*}$ represent the shell average in the spectral space, the Fourier coefficient, and the complex conjugate of the Fourier coefficients, respectively.
In Fig.\ref{fig:LESspec01}(a), $E_K(k,t)$ of run 002 piles up slightly at $k \geq 30$, suggesting the dissipation by the SGS viscosity $\nu_{{}_{SGS}}$ being too small.
The pile-up can be also seen $E_M(k,t)$ of run 002 in Fig.\ref{fig:LESspec01}(b).
These pile-ups are responsible for the large overestimate of $Q$ and $J$ in Fig.\ref{fig:LESstats01}(c) and (d).
The parameter set 000 gives a reasonable reproduction of $E_K(k,t)$ and $E_M(k,t)$, and other parameter sets, of which $C_{\nu}$ and $C_{\eta}$ are larger than those of 002, give a steeper decay of the spectra at $k \geq 30$.

In Fig.\ref{fig:LESiso00}, isosurfaces of the enstrophy density $q$ and the current density $I$ of (a) runs 000 and (b) 003 are shown at $t=1$.
The variable $q$ is drawn in blue-green colors and $I$ in gray throughout this article.
In each panel, thresholds of $q$ and $I$ are changed depending on its mean ($m$) and deviation ($\sigma$) values so that we can see spatial structures of the two quantities clearly.
Although we expect some specific structures such as sheets and tubes in the isosurfaces of $q$ and $I$, the isosurfaces in (a) are cloudy and structures therein are not clear.
Detailed structures are difficult to distinguish from the cloudy isosurfaces even when we lower or raise the thresholds except for a few tubular vortices.
The isosurfaces of run 002, which is not shown here, does not show a clearer spatial structures than run 000.
The isosurfaces in (b) show clear vortex tubes and current sheets.
However, run 003 underestimates $E_M$ in Fig.\ref{fig:LESstats01}(b) too much, and the decay of the energy spectra at $k > 30$ is also steep in Fig.\ref{fig:LESspec01}.
Furthermore, the Smagorinsky constants of the parameter set 003 are very large, and equivalent to $C_{\nu} = 0.046$ (which has been given theoretically in Ref.\cite{Hamba2010}) and $\Delta \simeq 3.34 \Delta_0$.
This means that the essential resolution in the LES is three times ore more coarser than that in the parameter set 002.
This is why we need to make a compromise on the spatial structures and the other quantities.

\subsection{LES for $k_{max} > k_H$ with the new SGS model}
Now we move on to LES with the new SGS model (\ref{eq:SGS8}), with the parameter sets 100-109.
In the parameter sets 100-109, we set $C_{\eta} = C_{\nu} \times (5/7)$ as in Ref.\cite{Hamba2010}.
The parameters $\sigma_{{}_{H,1}}$ and/or $\sigma_{{}_{H,2}}$ are non-zero in the parameter sets 100-109.
These two parameters are new in the SGS model (\ref{eq:SGS8}).

Figure \ref{fig:LESstats02} is (a) $E_K$, (b) $E_M$, (c) $Q$, and (d) $J$ of runs 100-109.
In Fig.\ref{fig:LESstats02}(a), all the LES give the time evolution of $E_K$ very close to that DNS GS data.
Though the parameter sets 102 and 109 give relatively large deviation of $E_K$ from that of DNS GS data, the deviation is smaller than those in Fig.\ref{fig:LESstats01}.
In Fig.\ref{fig:LESstats02}(b), $E_M$s of all the LES are smaller than that of DNS GS data.
The Smagorinsky constant $C_{\eta}$ in the parameter sets 100 and 102 are $5/7$-times of $C_{\eta}$ in the parameter sets 000 and 002, respectively.
In spite of the smaller $C_{\eta}$, $E_M$ in runs 100 and 102 underestimate $E_M$ considerably in contrast to runs 000 and 002.
Although the new model causes underestimation of $E_M$, $Q$ and $J$ are reproduced more precisely by the new model.
In Fig.\ref{fig:LESstats02}(c) and (d), the LES either overestimate or underestimate $Q$ and $J$, respectively, depending on the parameters $C_{\eta}$, $C_{\nu}$, $\sigma_{{}_{H,1}}$ and $\sigma_{{}_{H,2}}$.

We can compare Figs.\ref{fig:LESstats01} and \ref{fig:LESstats02} to figure out a difference due to these new parameters.
For example, parameter sets 003, 103, and 109 have the same $C_{\nu} = 0.23$ and different $C_{\eta}$ and $\sigma_{{}_{H,1}}$.
We also compare the results of runs 100 to 000, and those of runs 101 to 001, respectively.
The deviation of $Q$, and $J$ of runs 100, 101, 103, and 109 from the DNS GS data are smaller than those of runs 000 and 003, primarily because of the smaller $C_{\eta}$.
The parameter sets 101 and 105 appear to give the best fitting of $Q$ and $J$ among the parameters 100-109, although the decay rate of $Q$ at $t > 0.5$ appears too slow in comparison to that of the GS component of DNS.
These comparisons suggest that the new SGS model can give a better compromise between the reproduction of the GS DNS data of $E_K$, $Q$ and $J$ in exchange for the underestimation of $E_M$.

In Fig.\ref{fig:LESspec02}, the energy spectra (a)$E_K(k,t)$ and (b)$E_M(k,t)$ at $t=1$ of runs 100-104 are shown, while (c) and (d) are for the two spectra of runs 105-109, respectively.
The two spectra $E_K(k,t)$ and $E_M(k,t)$ of run 100 collapse with those of runs of 106 and 108, respectively, when they are plotted together.
The two spectra of runs 101 and 107 also collapse with those of runs 106 and 108 except a small difference at $k > 30$.
The energy spectra of the three runs 100, 106, and 108 appear going along $E_K(k,t)$ of DNS better than those of 002, without showing pile-up as in run 000 in Fig.\ref{fig:LESspec01}.
The energy spectra $E_K(k,t)$ and $E_M(k,t)$ of runs 102, 103, and 109 also collapse each other, showing a steeper decay at $k > 30$ of $E_M(k,t)$ than that of runs 100, 101, 106, 107, and 108.
The steepest decay of $E_M(k,t)$ is obtained by the parameter sets 104 and 105, in which $\sigma_{{}_{H,2}}=1$ provides a strong hyper-diffusivity.
Among the parameter sets 100-109, the parameter set 108 and 109 have the largest $\sigma_{{}_{H,1}}=8$.

We pay attention to run 108 because this parameter set gives not only a good energy spectrum profiles in Fig.\ref{fig:LESspec02}, but also the best fitting to $E_K$ and $E_M$ in Fig.\ref{fig:LESstats02} (a) and (b).
Although $Q$ and $J$ are overestimated in Fig.\ref{fig:LESstats02} (c) and (d), respectively, the difference from the DNS GS data is much smaller in comparison to those in the runs 000 and 002.
Additionally, the two spectra $E_K(k,t)$ and $E_M(k,t)$ of run 108 can be compared to those of run 104 because the parameters of the two runs are the same but $\sigma_{{}_{H,1}}$.
Obviously, $E_K(k,t)$ and $E_M(k,t)$ of run 108 in Fig.\ref{fig:LESspec02} fit well with the spectra of DNS better than those of run 104.
The comparison shows that the new term with the coefficient $\sigma_{{}_{H,1}}$ can improve the spectral nature of the LES.
Based on the observation in the above, we hereafter focus on runs 105-109.

In order to investigate the LES results further, we see spatial structures of the enstrophy density $q=\omega_i \omega_i/2$ and the current density $I=j_i j_i/2$.
Two aspects of the two quantities are of interest here.
The first aspect is that high-intensity regions of $q$ and $I$ should be well separated from each other in Hall MHD turbulence while they often stay together in single-fluid MHD turbulence\cite{Miura2014a}. 
The Hall term, representing a separation effect of the ion and electron motions, changes the frozen-in condition between the velocity field and the magnetic field, and, consequently, the spatial distributions of high-intensity regions of $q$ and $I$, too.
Thus we can see an effect of the Hall term visually through the separation of $q$ and $I$.
The second aspect is that the vortex structures are transformed from sheets to tubes as the consequence of the introduction of the Hall term, and the structure transition can be observed not only at the scale of the ion skin depth but also at a larger scale.
Since these two aspects are essential change from single-fluid to Hall-MHD turbulence, we require LES of Hall MHD turbulence to reproduce these two aspects.

Fig.\ref{fig:LESiso01} is isosurfaces of $q$ and $I$ of runs (a)105, (b)106, (c)107, and (d)108 at $t=1$.
In Fig.\ref{fig:LESiso01}(a), the isosurfaces of $q$ and $I$ are unclear, their shapes being neither sheets nor tubes, and distinguishing isosurfaces of $q$ from those of $I$ is not very simple.
In contrast, vortex tubes and sheets are clearly observed, and isosurfaces of $q$ and $I$ can be distinguished easily in Fig.\ref{fig:LESiso01}(b)-(d).
The visualizations in Fig.\ref{fig:LESiso01}(b)-(d) suffice for the two aspects described in the previous paragraph. 
Among the three panels, Fig.\ref{fig:LESiso01}(d) shows the clearest vortex tubes/sheets and current sheets, and looks the most favorable.
Here we remark on how we give the thresholds of $q$ and $I$.
In Fig.\ref{fig:LESiso01}(a), the thresholds of the two quantities are given by $m + 4\sigma$.
In Fig.\ref{fig:LESiso01}(b) and (d), the thresholds are given by $m + 3\sigma$, whereas they are $m + 2\sigma$ in (c).
The threshold $m + 2\sigma$ gives about $5\%$ of the volume if the scalar field (either $q$ or $I$) has the Gaussian distribution.
The fact that we need to change the thresholds in the visualizations (a)-(d) suggests that distributions of $q$ and $I$ are qualitatively changed.

In order to clarify the qualitative change, we observe the Probability Density Functions (PDFs) of the vorticity and current density of runs 105-108.
In Fig.\ref{fig:LESpdf01}, PDF of the first component of (a) the vorticity and (b) the current density at $t=1$ are shown.
The two thick lines represent PDFs by DNS, operated by the low-pass filter of $k_{cut}=42$ and $21$.
In Fig.\ref{fig:LESpdf01}(a), the PDFs of all of the four runs are scattered near the plot of DNS GS data with $k_{cut} = 42$, and none of them are plotted near the DNS GS data with $k_{cut}=21$.
These plots show that the four runs reproduce vorticity distribution fairly well for the resolution of $N^3=128^3$.
In Fig.\ref{fig:LESpdf01}(b), in contrast to (a), the PDFs of the current density of the four runs do not collapse.
While the plots of the PDFs of the runs 106, 107, and 108 collapse near the PDF of DNS GS data of $k_{cut}=42$, the PDF of the run 105 collapses with DNS GS data of $k_{cut}=21$.
The latter indicates that the run 105 has the resolution comparable to $N^3=64^3$ grid points.
This observation can be interesting because a nature of the two new terms in the Hall SGS model (\ref{eq:SGS8}), multiplied by $\sigma_{{}_{H,1}}$ and $\sigma_{{}_{H,2}}$ appears very well.
The two Smagorinsky constants $C_{\nu}$ and $C_{\eta}$ in the run 105 are the same as those of the run 108.
A difference in the parameters between the runs 105 and 107 is in $\sigma_{{}_{H,1}}$ and $\sigma_{{}_{H,2}}$.

Since $\sigma_{{}_{H,2}}=1$ in the run 105, the fourth-order hyper-diffusivity can reduce the effective numerical resolution of the magnetic field from $N^3=128^3$ to $N^3=64^3$.
However, we set $\sigma_{{}_{H,2}}=1$ in the run 108, too.
Nevertheless, the PDF of $j_1$ of run 108 is scattered near the PDF of DNS GS data of $k_{cut}=42$, without showing the reduction of the effective numerical resolution.
This indicates that the second term in the right-hand-side (RHS) of eq.(\ref{eq:SGS8}), multiplied with $\sigma_{{}_{H,1}}=8$ in the run 108, keeps the effective numerical resolution against the fourth-order hyper-diffusivity in eq.(\ref{eq:SGS8}).
It should be remembered here that $E_M(k,t)$ of run 108 fits better to that of DNS in comparison to run 105 as we have seen in Fig.\ref{fig:LESspec02}, indicating that the second term in the RHS of eq.(\ref{eq:SGS8}) can improve the spectra property as well.

From these numerical results, we consider that our new SGS model in eq.({eq:SGS8}) improves spectral and statistical natures of LES of Hall MHD turbulence well, especially when the parameter set 108 is adopted.
The Smagorinsky coefficient $C_{\nu}=0.092$ for the filter width $\Delta=\Delta_0$ is equivalent to $C_{\nu} = 0.046$ and $\Delta=1.68 \Delta_0$.
This means that the numerical resolution in the LES with the parameter set 108 is two times finer than that with the parameter set 003.

\subsection{LES for $k_{max} \simeq k_H$ }
We have studied LES of $k_{max} \simeq 2 k_H > k_H$ in the previous subsection.
However, we wish to set $k_{max} \leq k_H$ in an LES because resolving scales $k_{max} \simeq 2 k_H$ is still expensive for applications of the Hall MHD equations.
Though making a numerical resolution too coarse in an LES can change nonlinear dynamics in GS qualitatively from the original governing equation, it is still worth examining the marginal region $k_{max} \simeq k_H$.
In order to study influences of the ion skin depth scale on LES of Hall MHD turbulence, LES with the parameter sets 105-108 are carried out, firstly with the number of grid points $N^3=128^3$ and the Hall parameter $\epsilon_H=0.025$ ($k_{max}=42$ and $k_{H}=40$), and secondly with the number of grid points $N^3=64^3$ and the Hall parameter $\epsilon_H=0.05$ ($k_{max}=21$ and $k_{H}=20$).
In both cases, the relation between the maximum resolution of LES and the ion skin depth is $k_{max} > k_H$.

In Fig.\ref{fig:LESstats03}, time evolution of (a) $E_K$, (b) $E_M$, (c) $Q$, and (d) $J$ for LES with the parameter sets 105-108, $N^3=128^3$, and $\epsilon_H{=0.05}$ are shown.
DNS GS data is also for $\epsilon_H = 0.025$.
The relation between the LES and DNS results are quite similar to that in Fig.\ref{fig:LESstats02}: the LES predict $E_K$ fairly well in (a), underestimate $E_M$ in (b), moderately reproduce $Q$ in (c) and $J$ in (d).
In Fig.\ref{fig:LESspec03}, (a) $E_K(k,t)$ and (b) $E_M(k,t)$ at $t=1$ of runs 105-108 are shown.
Again, the energy spectra in the figure are quite similar to those in Fig.\ref{fig:LESspec02}.
These observations indicate that we can apply our SGS model for LES of $k_{max} \simeq k_H$.
We recall that the LES results of single-fluid MHD turbulence with the parameter sets 000-004 show a large deviation from the DNS result at $k=1$, the largest scale, in Ref.\cite{Miura2016}.
Since single-fluid MHD can be understood as the limit of $\epsilon_H \rightarrow 0$ ($k_H \rightarrow \infty$ equivalently), the LES results are considered to be worse for $k_H > k_{cut}$.

In Fig.\ref{fig:LESstats04}, time evolution of (a) $E_K$, (b) $E_M$, (c) $Q$, and (d) $J$ for LES with the parameter sets 105-108 with $N^3=64^3$ and $\epsilon_H=0.05$ ($k_{max}=21$ and $k_H=20$) are shown, together with the DNS GS data of $k_{cut}=21$.
While the LES give time evolution of $E_K$ in (a) very close to the DNS GS data with $k_{cut}=21$, the LES show a clear tendency of overestimation of $E_K$.
The overestimation is clearer in (c) on $Q$.
On the other hand, time evolution of $E_M$ in (b), especially of runs 106 and 108 which collapse each other completely, is very close to $E_M$ of the DNS GS data in the initial stage of the time evolution at $t \leq 0.5$ (the time of the peak of $Q$ and $J$).
At this point, the LES of $N^3=64^3$ give a better agreement with DNS GS data than the LES of $N^3=128^3$ and $\epsilon_H=0.05$.
At $t > 0.5$, however, $E_M$ of LES are not along the time evolution of that of the DNS GS data.
This shows a contrast with the time evolution of $E_M$ in Fig.\ref{fig:LESstats02}(b).
This separation of the $E_M$ in the LES from the DNS GS data may come from the lack of a sufficient number of Fourier coefficients rather than from the relative relation of $k_{cut}$ and $k_{H}$.

Fig.\ref{fig:LESspec04} is for (a)$E_K(k,t)$ and (b)$E_M(k,t)$.
A clear change in Fig.\ref{fig:LESspec04} from Fig.\ref{fig:LESspec02} is that the LES with $N^3=64^3$ apparently underestimate $E_M(k,t)$ except $k=1$.
Though the mean magnetic energy is well reproduced by the LES, the energy spectrum $E_M(k,t)$ in Fig.\ref{fig:LESspec04} may not be acceptable for some applications because of clear underestimation of $E_M(k,t)$ at $k > 1$.
Considering observations in Figs.\ref{fig:LESstats04} and \ref{fig:LESspec04} together, the Fourier coefficients of $21 \leq k \leq 42$ which are eliminated from a $N^3=64^3$ simulation are coupled closely with low-$k$ coefficients, and play significant roles in the reproduction of the spontaneous magnetic energy spectrum and long-term behavior of the mean magnetic energy.
This point should be studied further in relation with the cascade/inverse-cascade nature of turbulence in a future paper.

In summary, the results of LES in this section show that we can apply our new SGS model successfully to homogeneous and isotropic Hall MHD turbulence within the range of $k_H \simeq k_{cut}$, but the reliability of the LES can depend on the number of grid points.

\section{Concluding Remarks}
We have developed a new sub-grid-scale model for a large eddy simulation of freely-decaying homogeneous and isotropic Hall Magnetohydrodynamic turbulence.
Natures of the new model have been studied on time evolution of the energies, enstrophy, total current, energy spectra of the kinetic and magnetic field, and PDFs of the vorticity and the current density.
Some combinations of the Smagorinsky constants have been assessed from the points of view of these quantities, with the help of the visualization of local structures as well as the probability density functions of the vorticity and current components.

LES of Hall MHD turbulence have shown that our new model improves spectral properties of LES, and provides reproduction of the DNS GS data.
While the previous model is also acceptable if we focus on the kinetic and magnetic energies, the new model achieves not only good spectral properties but also reproduces the kinetic energy, enstrophy, current, and local structure visualization in exchange for an underestimate of the magnetic energy.
Analysis for combinations of the numerical resolution and the Hall parameter shows that we can carry out LES not only for $k_{max} > k_H$ but also for $k_{max} \simeq k_H$. 
By the use of the new SGS, we can double an essential resolution of LES of Hall MHD turbulence in comparison to the previous SGS model. 
Our new SGS model enables numerical simulations of homogeneous and isotropic magnetohydrodynamic turbulence with Hall effects with a small computational cost, retaining the ion-electron separation effects by the Hall term in the grid scale.

This research was partially supported by JSPS KAKENHI Grant Number 17K05734 and MEXT KAKENHI Grant Number 15H02218, Japan. 
The numerical simulations were performed on the FUJITSU FX100 supercomputer {\it Plasma Simulator} of NIFS with the support and under the auspices of the NIFS Collaboration Research program (NIFS15KNSS053, NIFS15KNTS038), and partially on the NEC SX-ACE supercomputer of Tohoku University as well as on the FUJITSU Oakforest-PACS supercomputer of the University of Tokyo, being partially supported by ``Joint Usage/Research Center for Interdisciplinary Large-scale Information Infrastructures'' in Japan.

\clearpage

\section*{APPENDIX: MODELING THE HALL TERM}

In the appendix we describe the derivation of the model expression for the Hall term appearing in (\ref{eq:SGS8}). 
The SGS Hall electromotive force $\overline E_i^H$ defined in (\ref{eq:GSHallMHD8}) can be rewritten as

\begin{equation}
\overline E_i^H = {\epsilon _H}\frac{\partial }{{\partial {x_j}}}(\overline {{B_i}{B_j}}  - {\overline B_i}{\overline B_j}) - {\epsilon _H}\frac{1}{2}\frac{\partial }{{\partial {x_i}}}(\overline {{B_j}{B_j}}  - {\overline B_j}{\overline B_j}) .
\end{equation}
The second part on the right-hand side does not contribute to $ - \nabla  \times {{\bf{\overline E}}^H}$ in the induction equation given by (12). 
We need to model the magnetic field correlation $\overline {{B_i}{B_j}}  - {\overline B_i}{\overline B_j}$. 
Here, we use the Markovianized two-scale method for inhomogeneous turbulence modeling proposed by Yoshizawa \cite{yosh98}. 
For simplicity, we assume that $\overline {\overline f} = \overline f$ and $\overline {f'}  = 0$ where $f' = f - \overline f$ and $f$ is a quantity such as the velocity and the magnetic field. 
The magnetic field correlation is then written as $\overline {{B'_i}{B'_j}} $.

Using (\ref{eq:HallMHD4})-(\ref{eq:HallMHD6}) we can derive the equation for the fluctuating magnetic field ${B'_i}$ as follows:

\[
\frac{{\partial {B'_i}}}{{\partial t}} + \frac{\partial }{{\partial {x_j}}}{{\overline u}_j}{B'_i} + \frac{\partial }{{\partial {x_j}}}({u'_j}{B'_i} - \overline {{u'_j}{B'_i}} ) - \eta \frac{{{\partial ^2}{B'_i}}}{{\partial {x_j}\partial {x_j}}}
\]

\begin{equation} \label{eq:a2}
 =  - {u'_j}\frac{{\partial {{\overline B}_i}}}{{\partial {x_j}}} + \frac{{\partial {u'_i}}}{{\partial {x_j}}}{{\overline B}_j} + {B'_j}\frac{{\partial {{\overline u}_i}}}{{\partial {x_j}}} + {\epsilon _H}\left( {\frac{{\partial {B'_i}}}{{\partial {x_j}}}{{\overline J}_j} + {J'_j}\frac{{\partial {{\overline B}_i}}}{{\partial {x_j}}} - \frac{{\partial {J'_i}}}{{\partial {x_j}}}{{\overline B}_j} - {B'_j}\frac{{\partial {{\overline J}_i}}}{{\partial {x_j}}}} \right) ,
\end{equation}
where some insignificant terms are omitted. Only the terms on the left-hand side remain for homogeneous isotropic turbulence, while the terms on the right-hand side represent the anisotropic and inhomogeneous effects caused by the mean field. We expand fluctuating quantities appearing in (\ref{eq:a2}) as

\begin{equation} \label{eq:a3}
{u'_i} = {u'_{0i}} + {u'_{1i}} +  \cdots ,\ \ \ 
{B'_i} = {B'_{0i}} + {B'_{1i}} +  \cdots ,\ \ \ 
{J'_i} = {J'_{0i}} + {J'_{1i}} +  \cdots ,
\end{equation}
where the 0th-order terms denote the homogeneous isotropic turbulent field and the higher-order terms reflect the anisotropic and inhomogeneous effects. We substitute (\ref{eq:a3}) into (\ref{eq:a2}) and solve the turbulent field iteratively. We approximate that the Green's function for the operator on the left-hand side of (\ref{eq:a2}) can be replaced by the turbulent time scale $\tau$ \cite{yosh98}. We then obtain the expression for ${B'_{1i}}$ as follows:

\[
{B'_{1i}} = \tau \left( { - {u'_{0j}}\frac{{\partial {{\overline B}_i}}}{{\partial {x_j}}} + \frac{{\partial {u'_{0i}}}}{{\partial {x_j}}}{{\overline B}_j} + {B'_{0j}}\frac{{\partial {{\overline u}_i}}}{{\partial {x_j}}}} \right) 
\]

\begin{equation} \label{eq:a4}
+ {\epsilon _H}\tau \left( {\frac{{\partial {B'_{0i}}}}{{\partial {x_j}}}{{\overline J}_j} + {J'_{0j}}\frac{{\partial {{\overline B}_i}}}{{\partial {x_j}}} - \frac{{\partial {J'_{0i}}}}{{\partial {x_j}}}{{\overline B}_j} - {B'_{0j}}\frac{{\partial {{\overline J}_i}}}{{\partial {x_j}}}} \right) .
\end{equation}

The magnetic field correlation is expanded as

\begin{equation} \label{eq:a5}
\overline {{B'_i}{B'_j}}  = \overline {{B'_{0i}}{B'_{0j}}}  + \overline {{B'_{0i}}{B'_{1j}}}  + \overline {{B'_{1i}}{B'_{0j}}}  +  \cdots .
\end{equation}
Substituting (\ref{eq:a4}) into (\ref{eq:a5}) and considering the isotropy of the 0th-order quantities, we have

\[
\overline {{B'_i}{B'_j}}  = \frac{1}{3}\overline {B'^2_{0k}} {\delta _{ij}} + \frac{\tau }{3}\overline {B'^2_{0k}} {\overline S_{ij}} - \frac{{{\epsilon _H}\tau }}{3}\overline {B'^2_{0k}} \left( {\frac{{\partial {{\overline J}_i}}}{{\partial {x_j}}} + \frac{{\partial {{\overline J}_j}}}{{\partial {x_i}}}} \right)
\]

\begin{equation} \label{eq:a6}
 - \frac{\tau }{3}\left( {\overline {{u'_{0k}}{B'_{0k}}}  - {\epsilon _H}\overline {{B'_{0k}}{J'_{0k}}} } \right)\left( {\frac{{\partial {{\overline B}_i}}}{{\partial {x_j}}} + \frac{{\partial {{\overline B}_j}}}{{\partial {x_i}}}} \right) .
\end{equation}
The cross helicity $\overline {{u'_{0k}}{B'_{0k}}}$ and the current helicity $\overline {{B'_{0k}}{J'_{0k}}}$ show non-zero values for special kind of MHD turbulence only, whereas the magnetic energy $\overline {B'^2_{0k}}$ always exists. 
Therefore, we keep the first three terms on the right-hand side of (\ref{eq:a6}). 
The magnetic field correlation also appears in the SGS Reynolds stress defined in (9), for which we adopt the eddy viscosity model given by (\ref{eq:SGS1}). 
Comparing the second term in (\ref{eq:a6}) with (\ref{eq:SGS1}), we can see that $(\tau /3)\overline {B'^2_{0k}}  \propto {\nu _{SGS}}$. 
Therefore, we obtain the model expression for $\overline {{B'_i}{B'_j}} $ as follows:

\begin{equation}
\overline {{B'_i}{B'_j}}  = \frac{1}{3}\overline {B'^2_k} {\delta _{ij}} + {\nu _{SGS}}{\sigma _{H,1}}{\overline S_{ij}} - {\epsilon _H}{\nu _{SGS}}{\sigma _{H,2}}\left( {\frac{{\partial {{\overline J}_i}}}{{\partial {x_j}}} + \frac{{\partial {{\overline J}_j}}}{{\partial {x_i}}}} \right) ,
\end{equation}
where ${\sigma _{H,1}}$ and ${\sigma _{H,2}}$ are non-dimensional model constants. 
Because the first term on the right-hand side does not contribute to $ - \nabla  \times {{\bf{\overline E}}^H}$, the model expression for the SGS Hall electromotive force can be written as follows:

\begin{equation}
\overline E_i^H = {\epsilon _H}\frac{\partial }{{\partial {x_j}}}({\nu _{SGS}}{\sigma _{H,1}}{\overline S_{ij}}) - \epsilon _H^2\frac{\partial }{{\partial {x_j}}}\left[ {{\nu _{SGS}}{\sigma _{H,2}}\left( {\frac{{\partial {{\overline J}_i}}}{{\partial {x_j}}} + \frac{{\partial {{\overline J}_j}}}{{\partial {x_i}}}} \right)} \right] .\end{equation}

\clearpage

\section*{Figure captions}

\begin{figure}
\begin{center}
\begin{minipage}{0.45\textwidth}
(a)\hspace*{0.9\textwidth}\;\; \\    
  \includegraphics[width=0.99\textwidth]{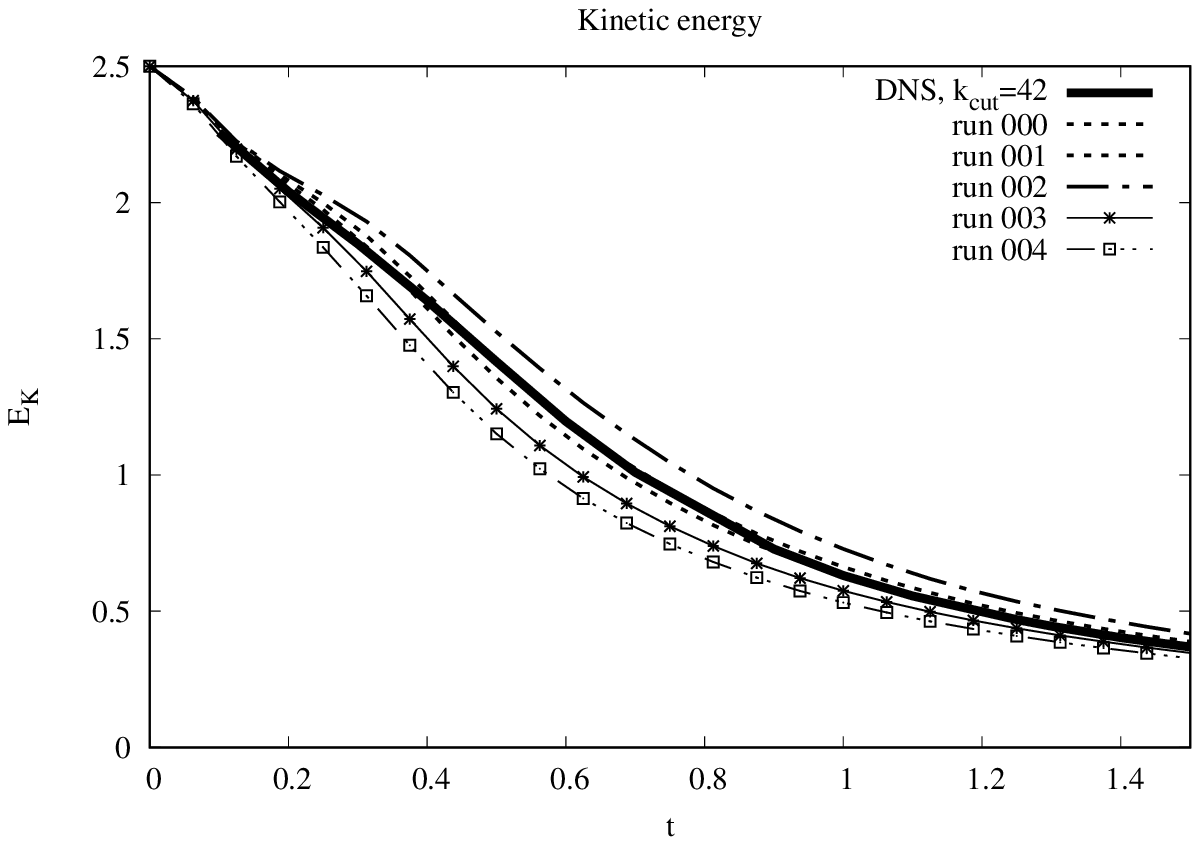} 
(b)\hspace*{0.9\textwidth}\;\; \\
  \includegraphics[width=0.99\textwidth]{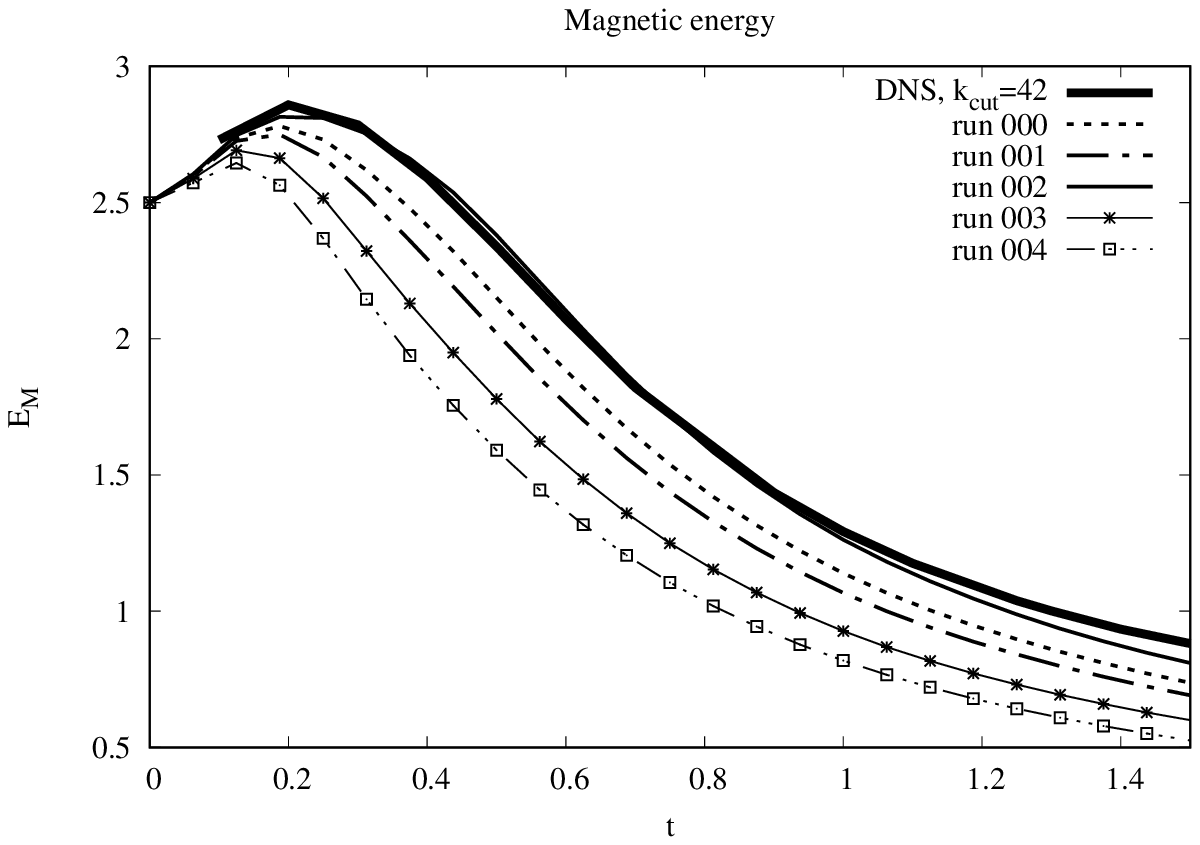}
\end{minipage}
\begin{minipage}{0.45\textwidth}
(c)\hspace*{0.9\textwidth}\;\;\\ 
  \includegraphics[width=0.99\textwidth]{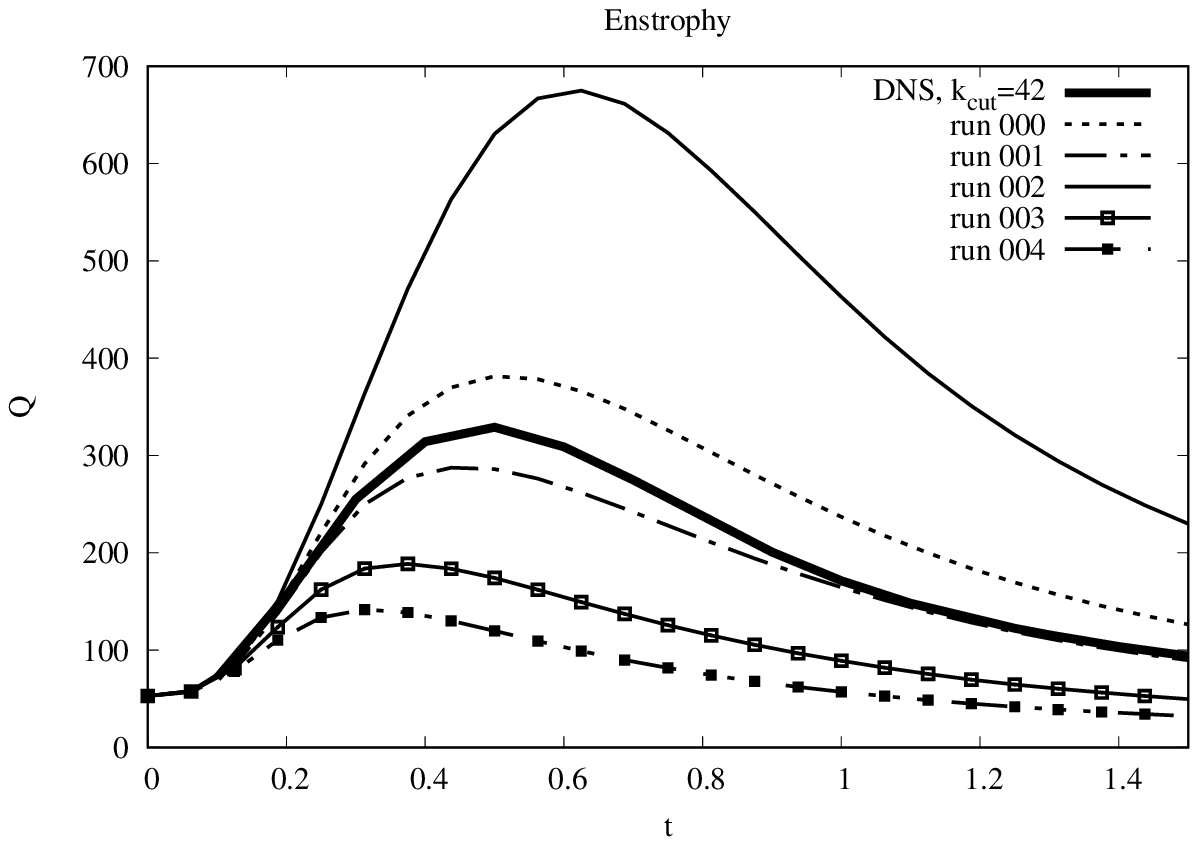}
(d)\hspace*{0.9\textwidth}\;\;\\ 
  \includegraphics[width=0.99\textwidth]{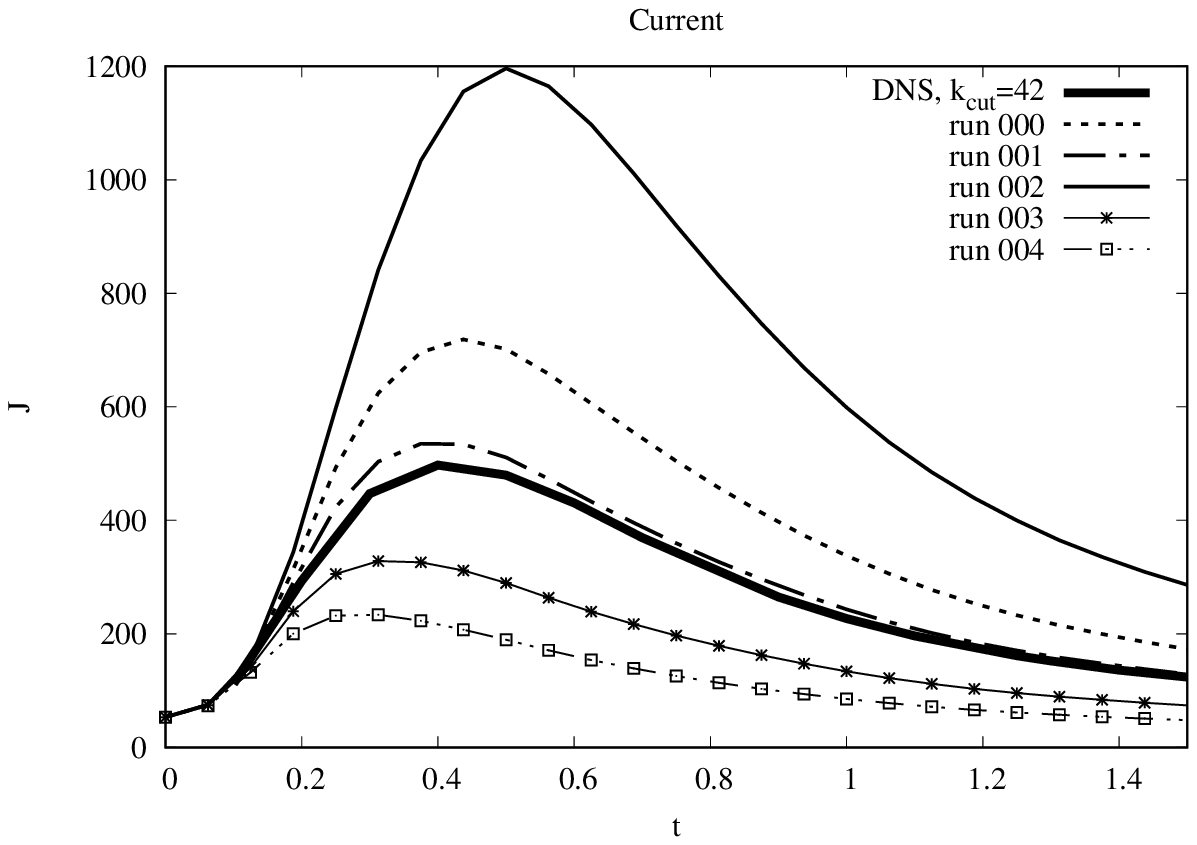}
\end{minipage}
$\;\;$ \\
$\;\;$ \\
\caption{\label{fig:LESstats01} Time evolution of (a) $E_K$, (b) $E_M$, (c) $Q$, and (d) $J$ of runs 000-004. }
\end{center}
\end{figure}

\begin{figure}
\begin{center}
(a)\hspace*{0.5\textwidth}\;\; \\    
  \includegraphics[width=0.55\textwidth]{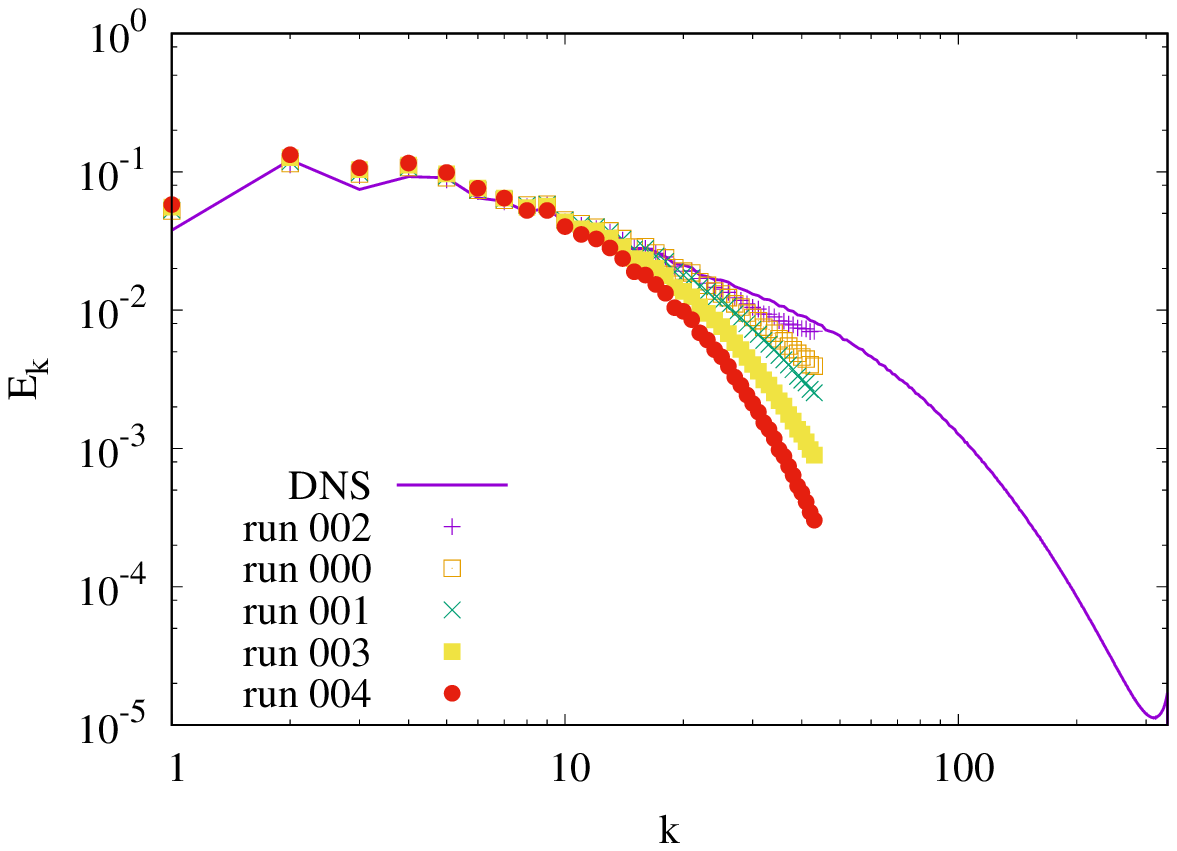}
(b)\hspace*{0.5\textwidth}\;\; \\    
  \includegraphics[width=0.55\textwidth]{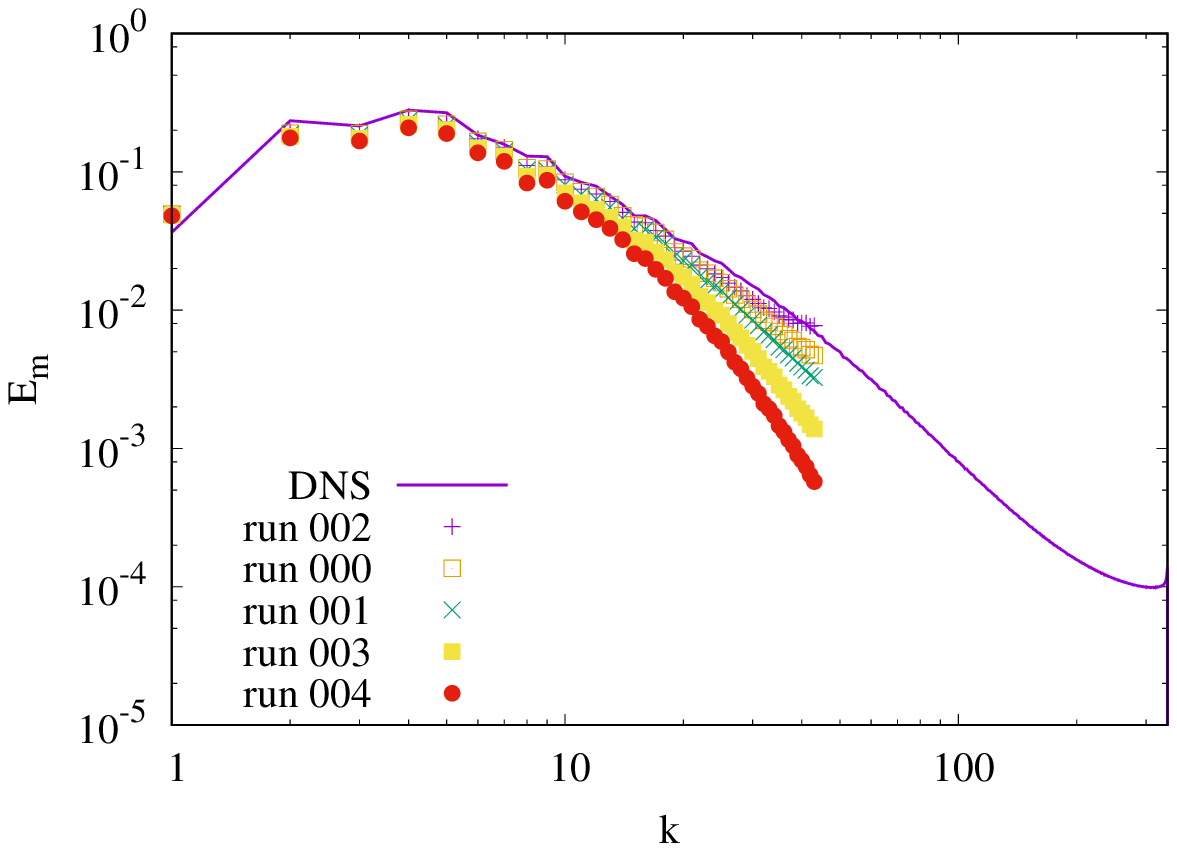}
$\;\;$ \\
$\;\;$ \\
\caption{\label{fig:LESspec01} Energy spectra (a) $E_K(k,t)$ and (b) $E_M(k,t)$ at $t=1$ of runs 000-004. }
\end{center}
\end{figure}

\begin{figure}
\begin{center}
\begin{minipage}{0.45\textwidth}
(a)\hspace*{0.8\textwidth}\;\; \\    
\includegraphics[width=0.99\textwidth]{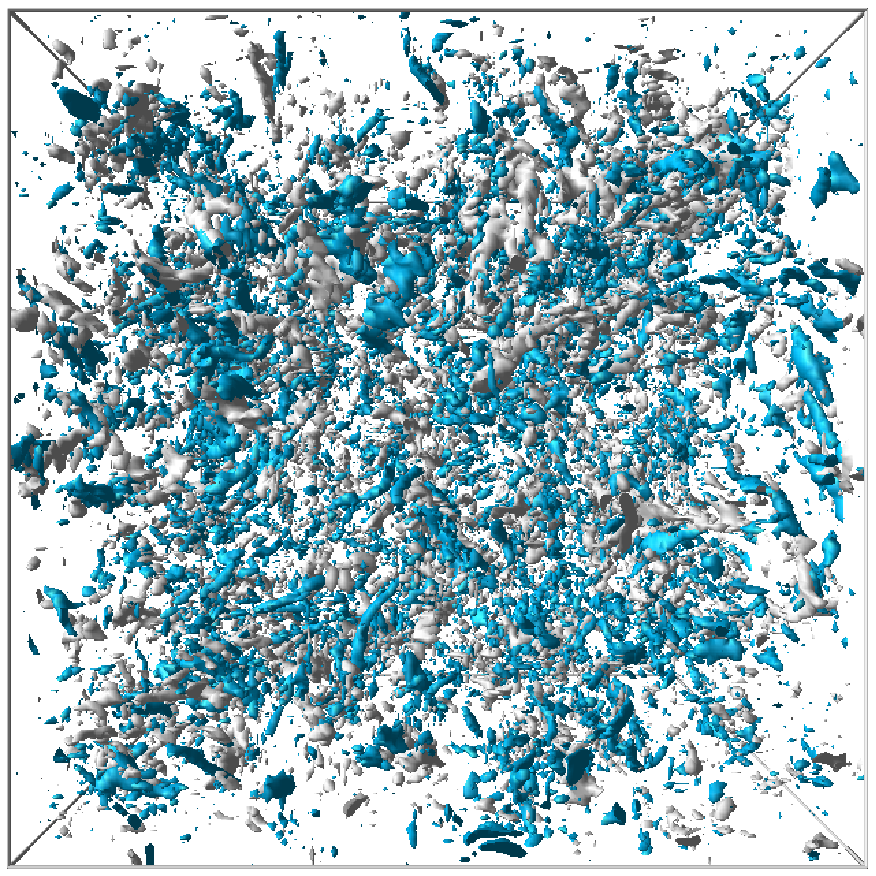} \\
(b)\hspace*{0.8\textwidth}\;\; \\    
\includegraphics[width=0.99\textwidth]{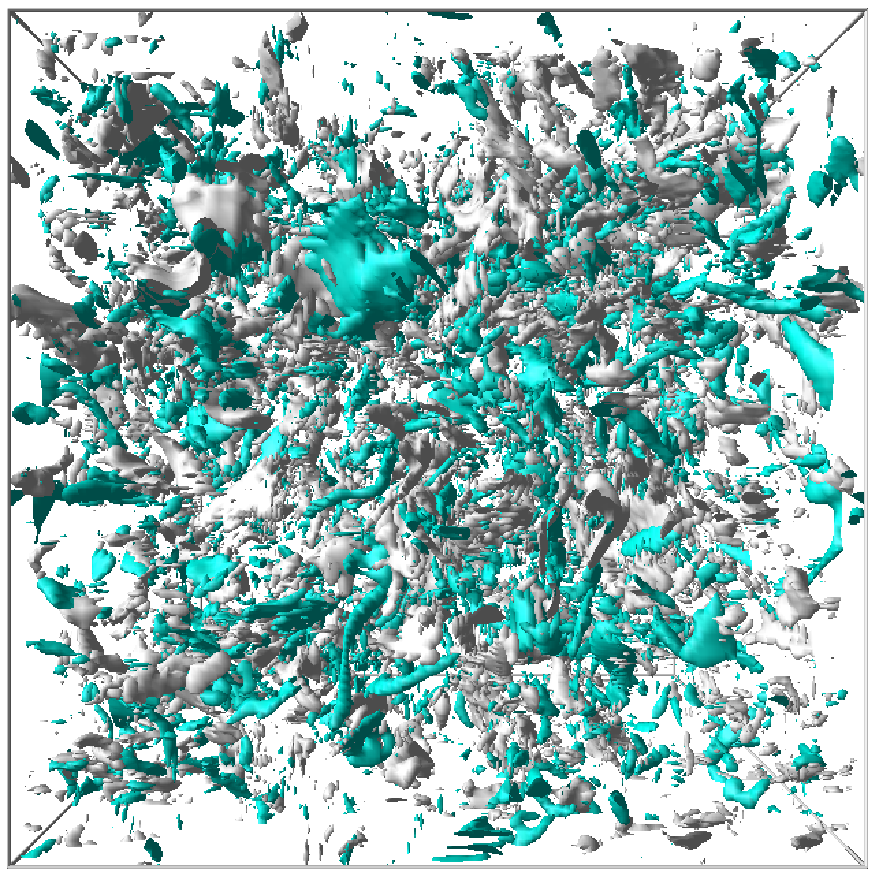}
\end{minipage}
$\;\;$ \\
$\;\;$ \\
\caption{\label{fig:LESiso00} Isosurfaces of $q$ (with blue-green colors) and $I$ (gray) at $t=1$ in LES of the runs (a) 000 and (b) 003.}
\end{center}
\end{figure}

\begin{figure}
\begin{center}
\begin{minipage}{0.45\textwidth}
(a)\hspace*{0.9\textwidth}\;\; \\    
  \includegraphics[width=0.99\textwidth]{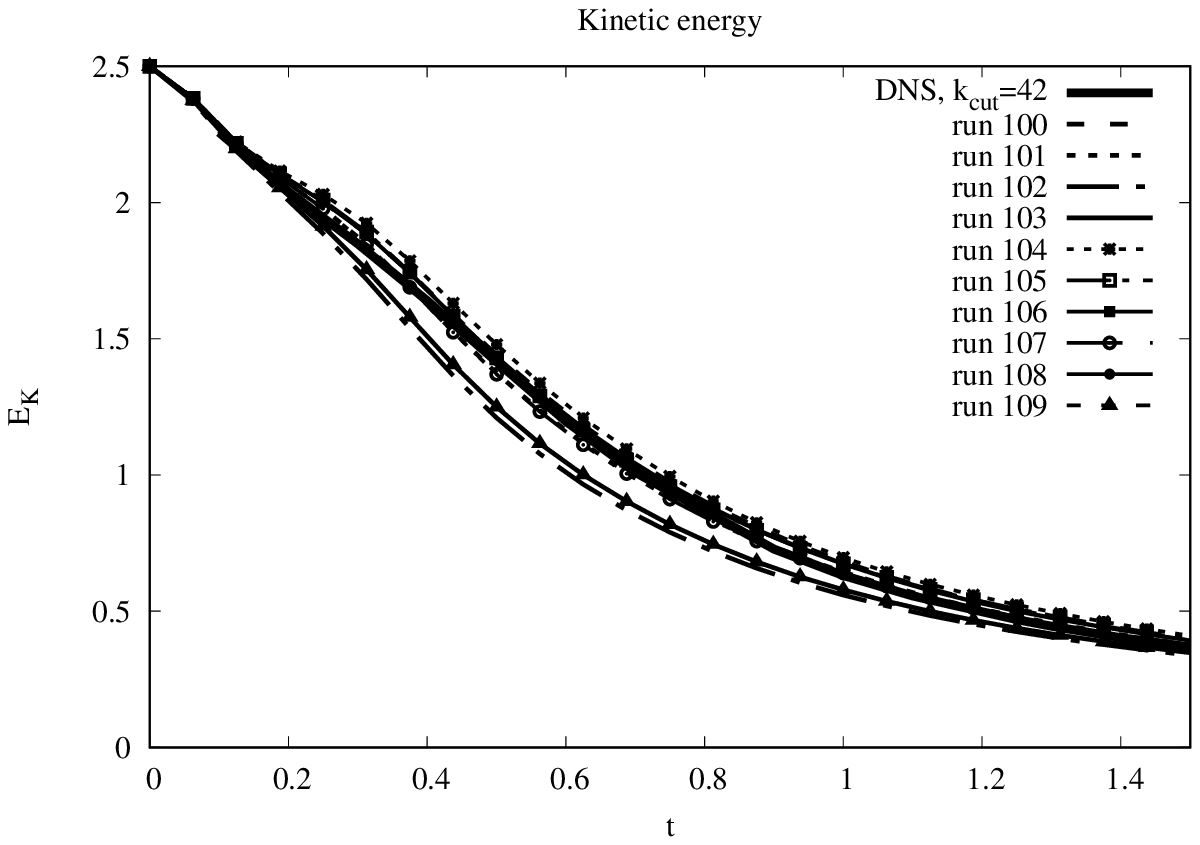}
(b)\hspace*{0.9\textwidth}\;\; \\    
  \includegraphics[width=0.99\textwidth]{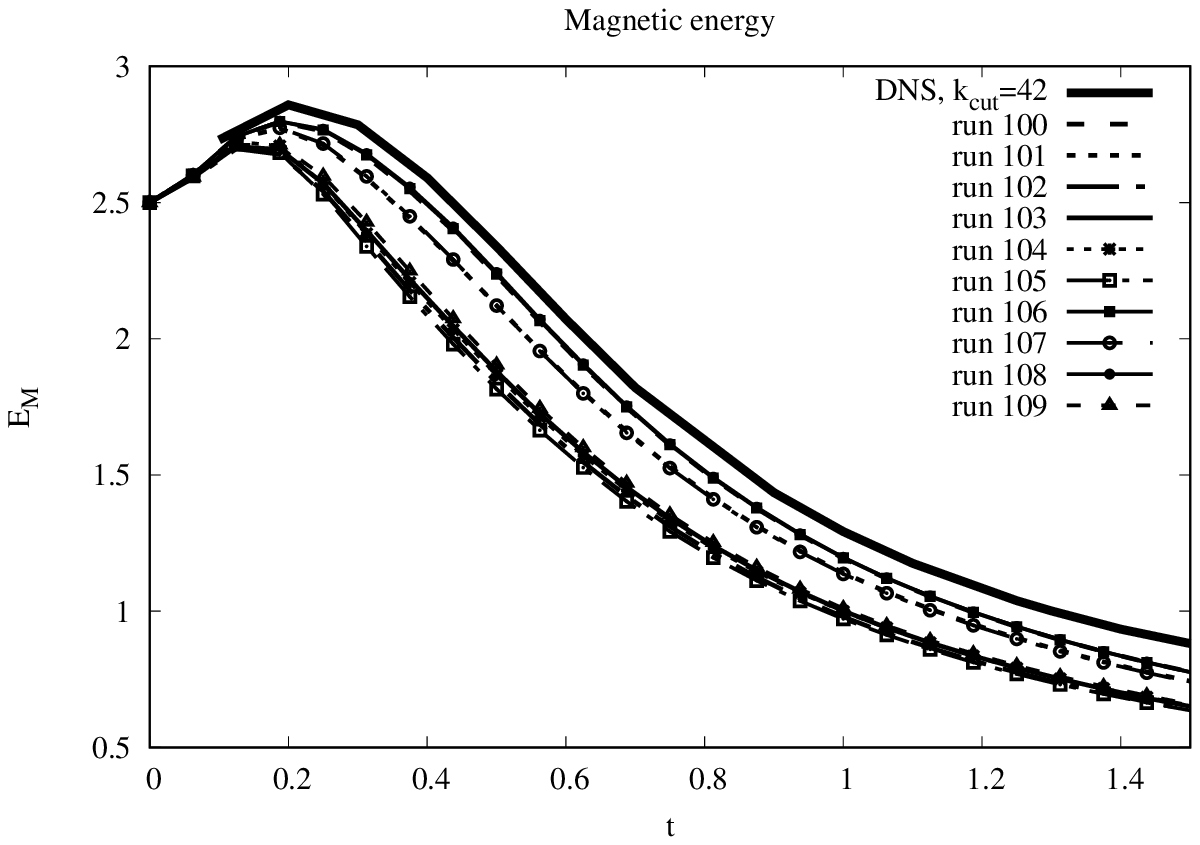}
\end{minipage}
\begin{minipage}{0.45\textwidth}
(c)\hspace*{0.9\textwidth}\;\; \\    
  \includegraphics[width=0.99\textwidth]{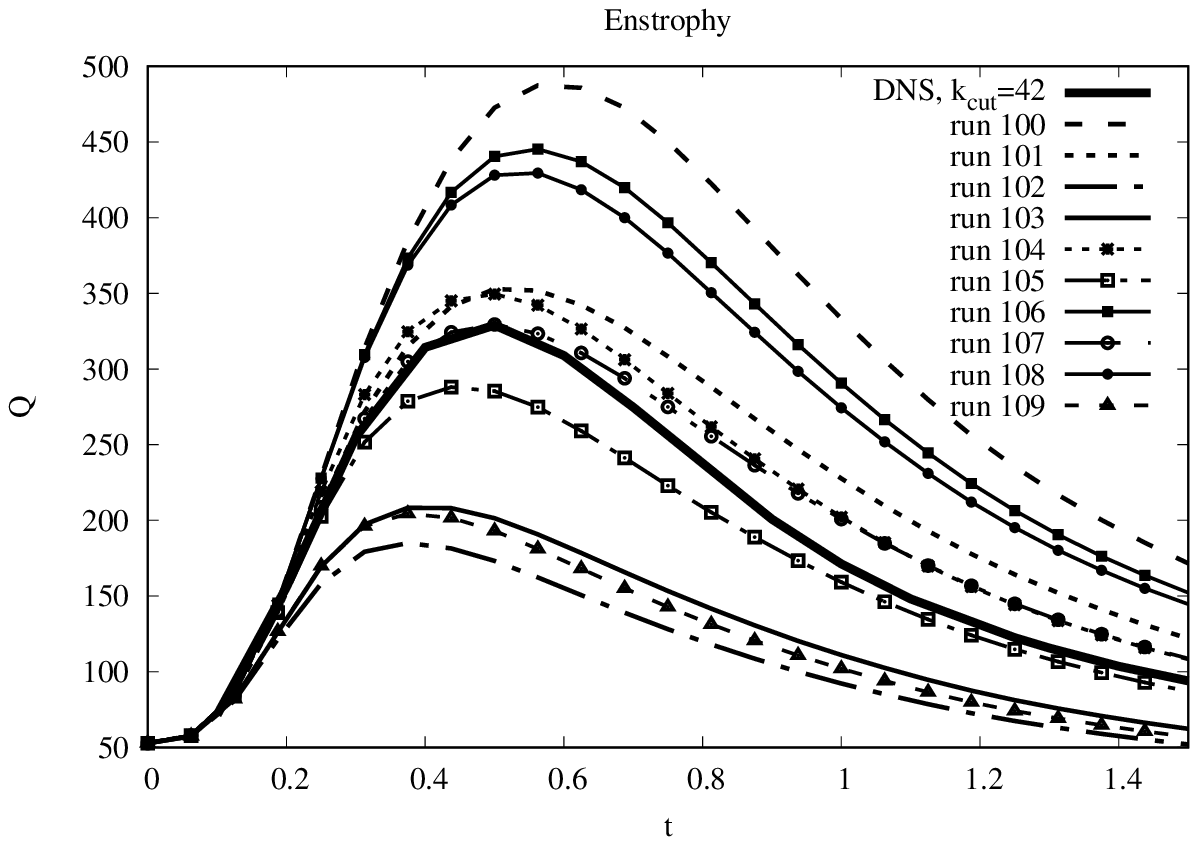}
(d)\hspace*{0.9\textwidth}\;\; \\    
  \includegraphics[width=0.99\textwidth]{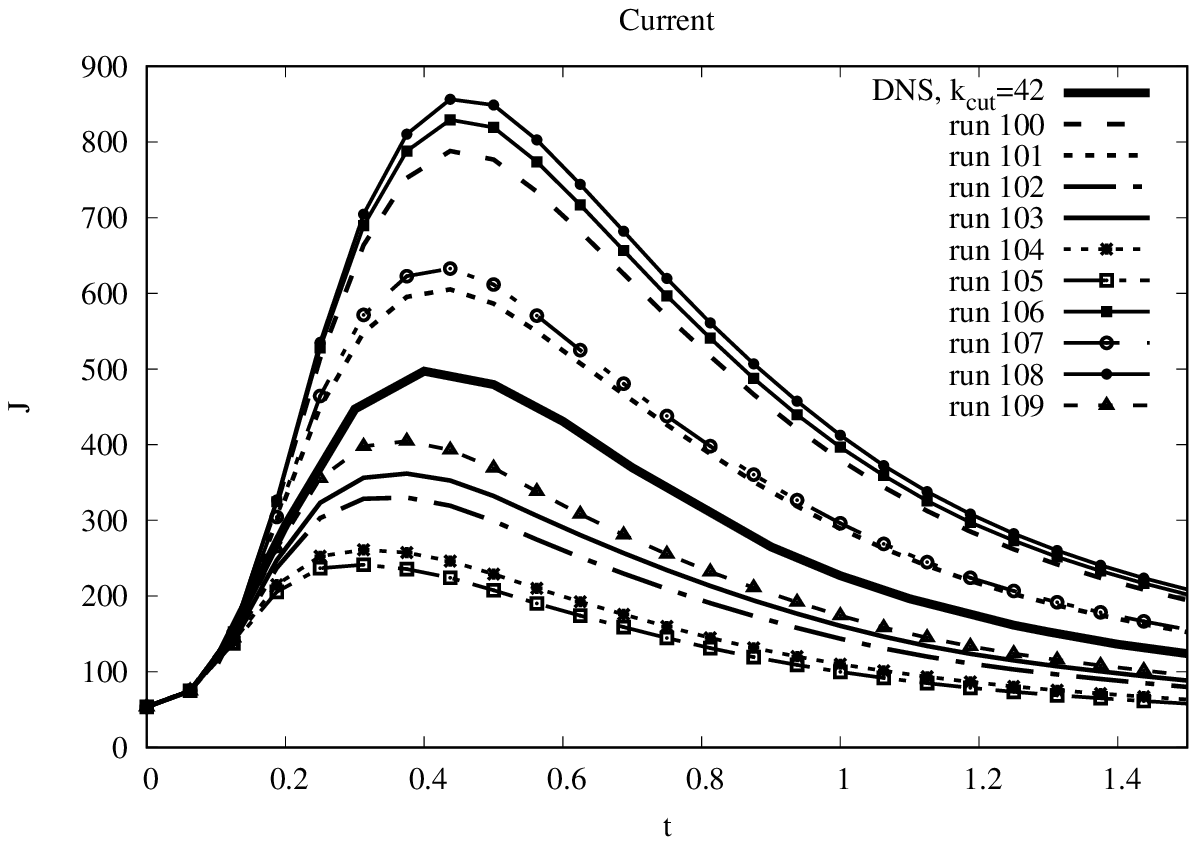}
\end{minipage}
$\;\;$ \\
$\;\;$ \\
\caption{\label{fig:LESstats02} Time evolution of (a) $E_K$, (b) $E_M$, (c) $J$, and (d) $Q$ of runs 100-109. }
\end{center}
\end{figure}

\begin{figure}
\begin{center}
\begin{minipage}{0.48\textwidth}
(a)\hspace*{0.9\textwidth}\;\; \\    
  \includegraphics[width=0.99\textwidth]{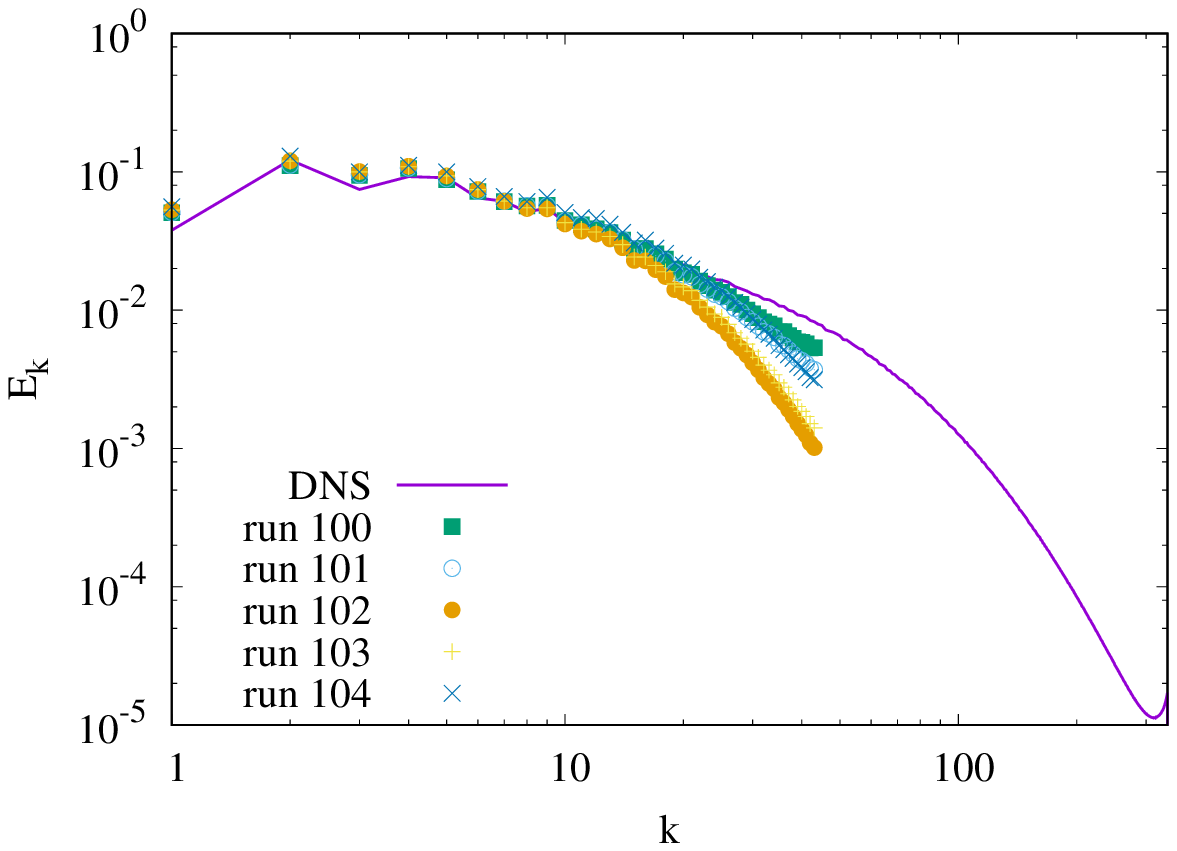}
(b)\hspace*{0.9\textwidth}\;\; \\    
  \includegraphics[width=0.99\textwidth]{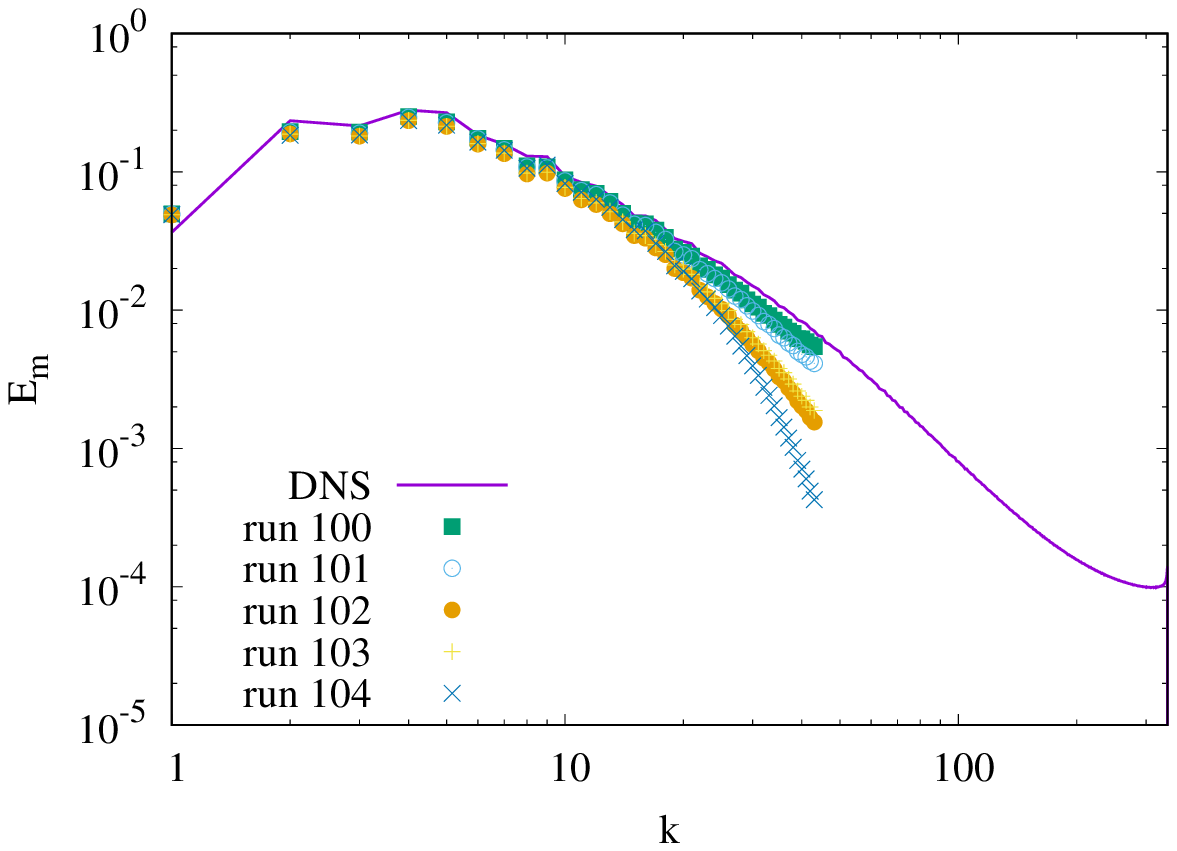}
\end{minipage}
\begin{minipage}{0.48\textwidth}
(c)\hspace*{0.9\textwidth}\;\; \\    
  \includegraphics[width=0.99\textwidth]{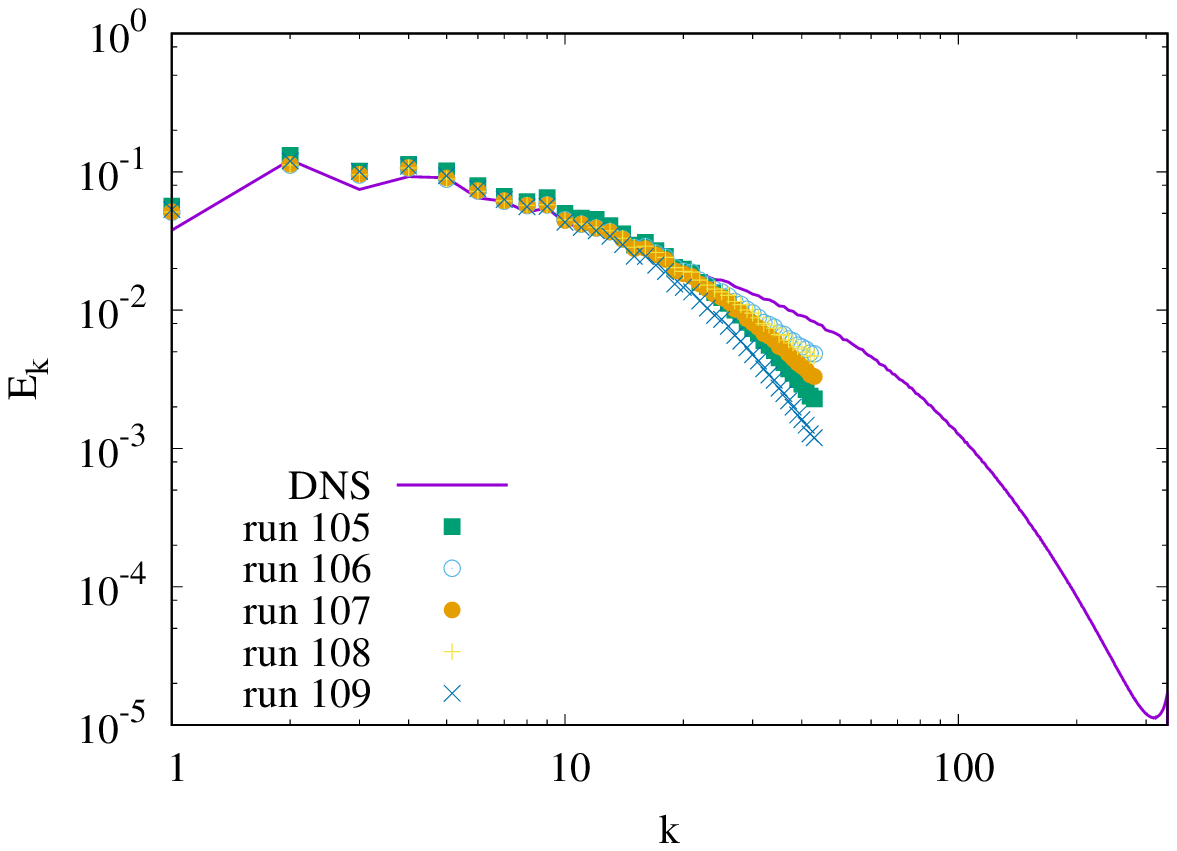}
(d)\hspace*{0.9\textwidth}\;\; \\    
  \includegraphics[width=0.99\textwidth]{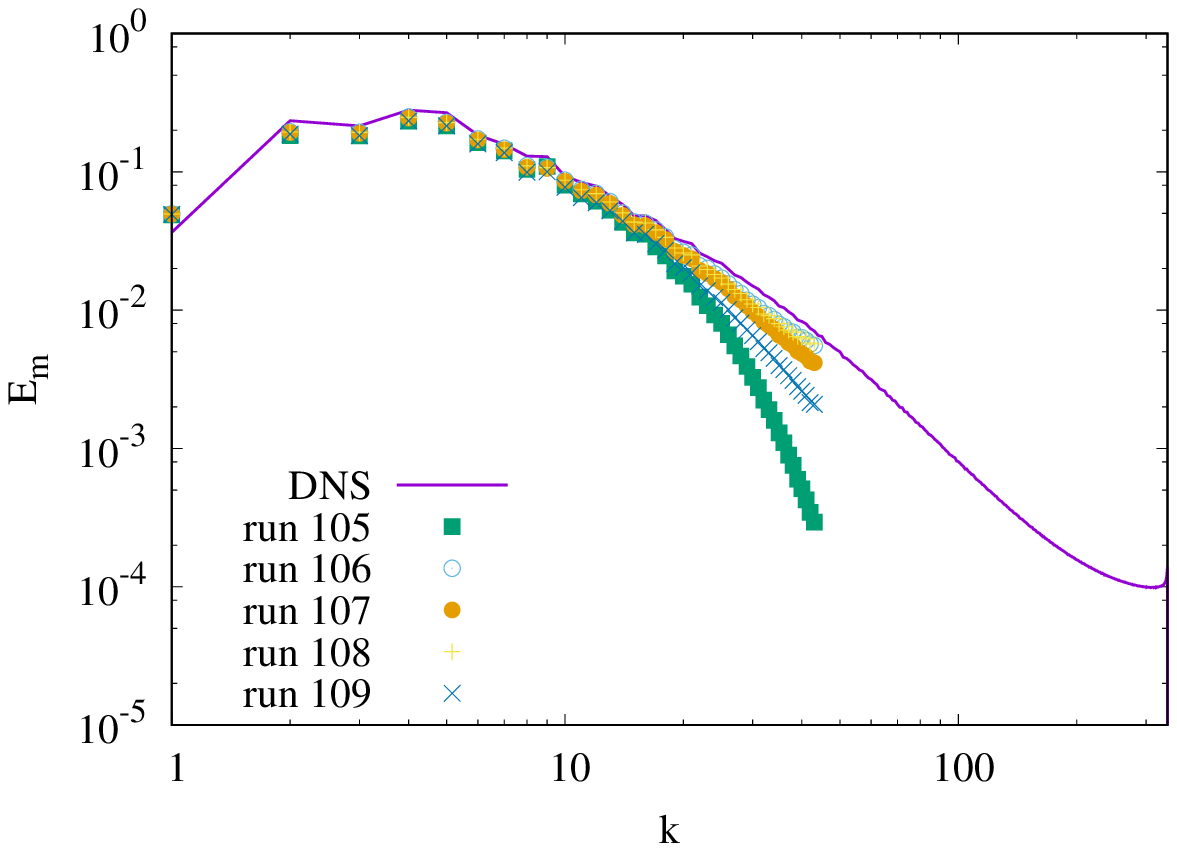}
\end{minipage}
$\;\;$ \\
$\;\;$ \\
\caption{\label{fig:LESspec02} Energy spectra (a) $E_K(k,t)$ and (b) $E_M(k,t)$ at $t=1$ of runs 100-104, and (c) $E_K(k,t)$ and (d) $E_M(k,t)$ at $t=1$ of runs 105-109. }
\end{center}
\end{figure}

\begin{figure}
\begin{center}
\begin{minipage}{0.45\textwidth}
(a)\hspace*{0.9\textwidth}\;\; \\    
  \includegraphics[width=0.99\textwidth]{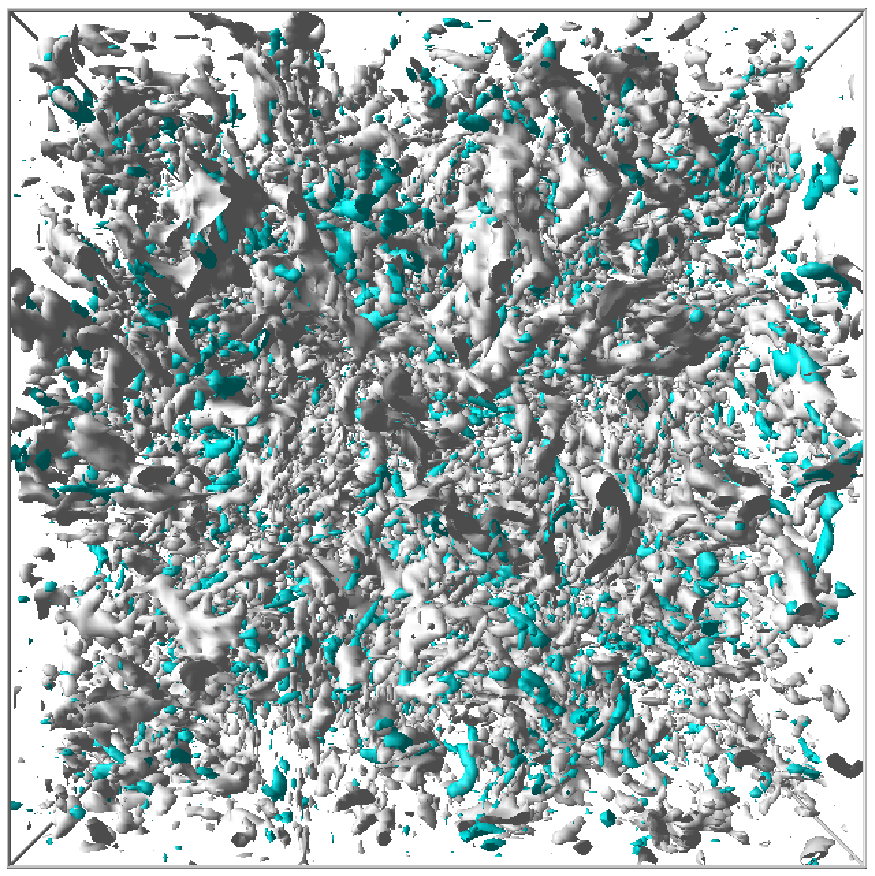}
(b)\hspace*{0.9\textwidth}\;\; \\    
  \includegraphics[width=0.99\textwidth]{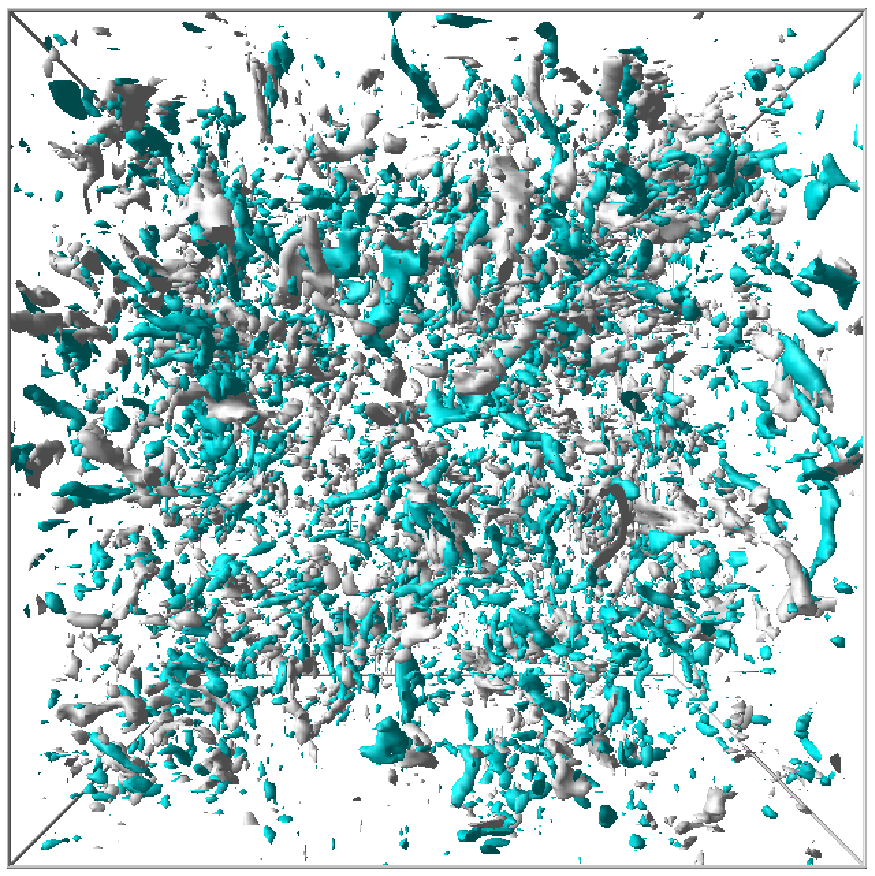}
\end{minipage}
\begin{minipage}{0.45\textwidth}
(c)\hspace*{0.9\textwidth}\;\; \\    
  \includegraphics[width=0.99\textwidth]{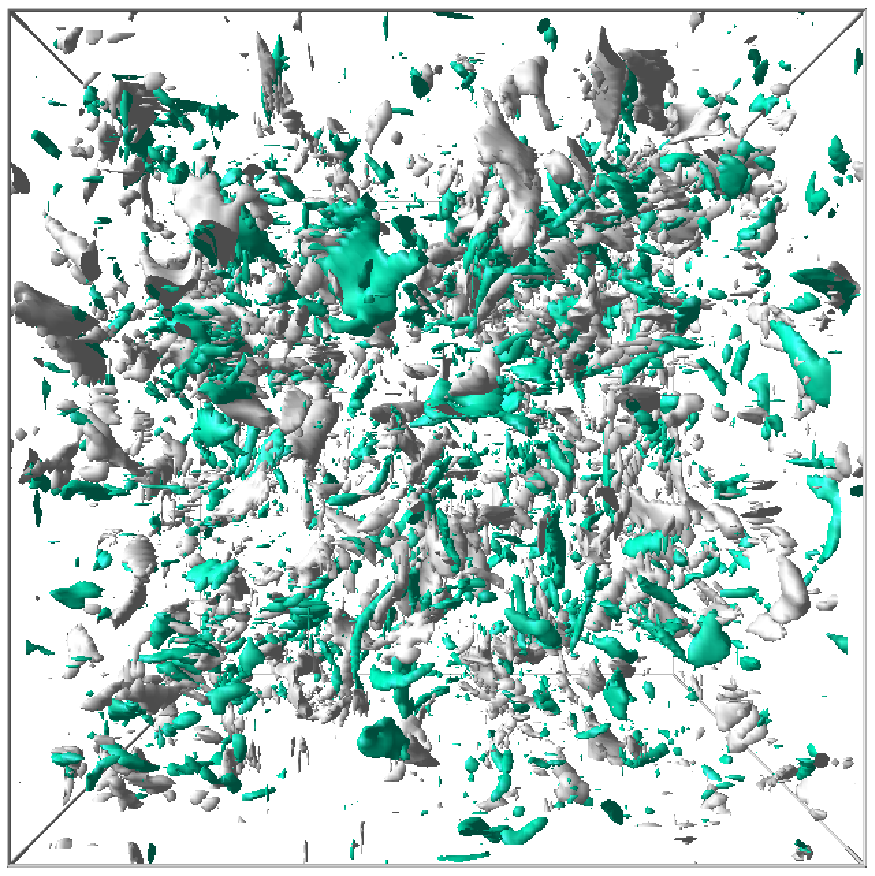}
(d)\hspace*{0.9\textwidth}\;\; \\    
  \includegraphics[width=0.99\textwidth]{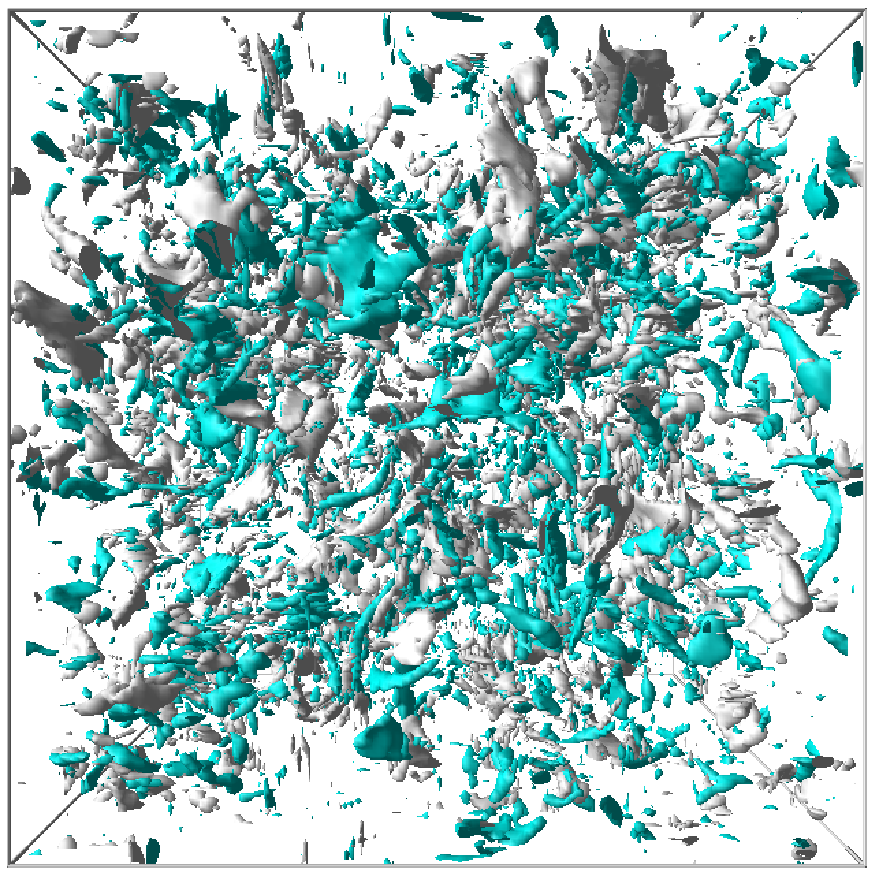}
\end{minipage}
$\;\;$ \\
$\;\;$ \\
\caption{\label{fig:LESiso01} Isosurfaces of $q$ (with blue-green colors) and $I$ (gray) at $t=1$ in LES of the runs (a) 105, (b) 106, (c) 107, and (d) 108.}
\end{center}
\end{figure}

\begin{figure}
\begin{center}
(a)\hspace*{0.5\textwidth}\;\; \\    
  \includegraphics[width=0.55\textwidth]{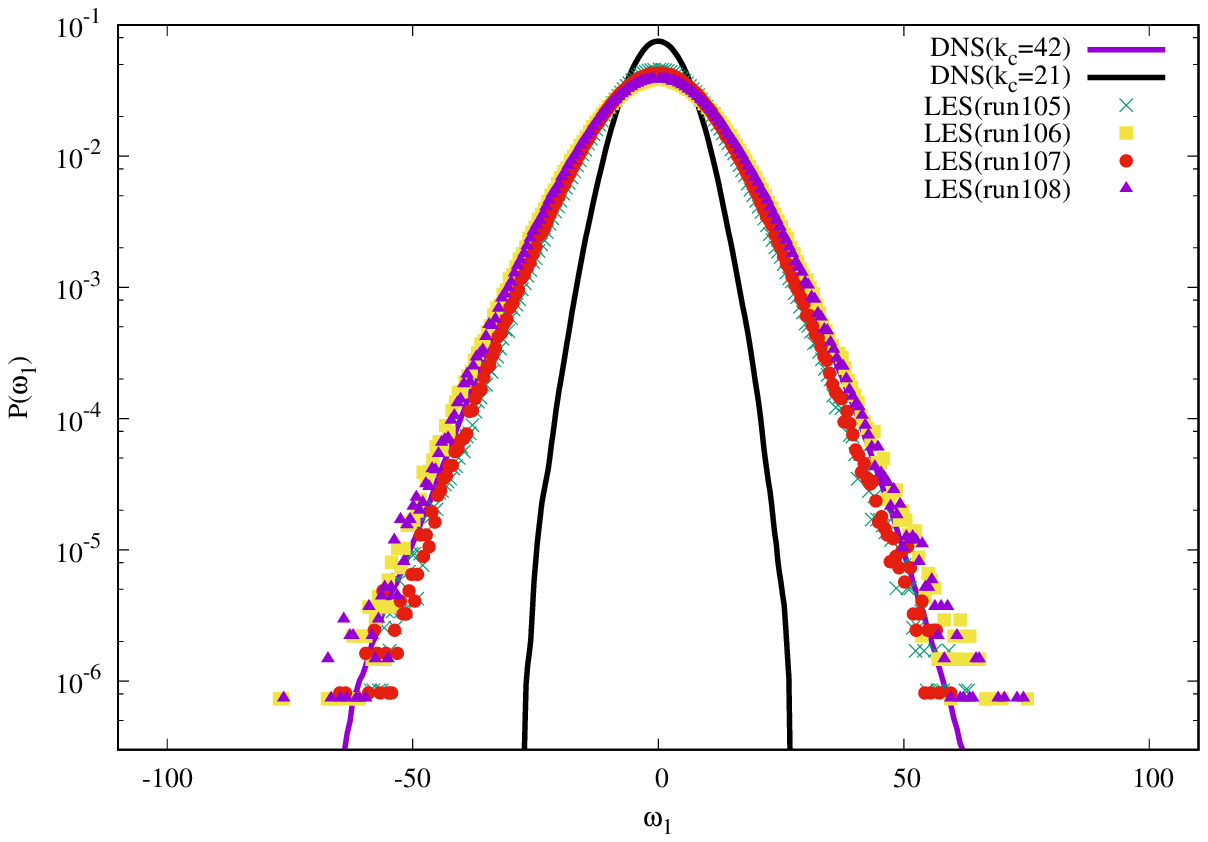}
(b)\hspace*{0.5\textwidth}\;\; \\    
  \includegraphics[width=0.55\textwidth]{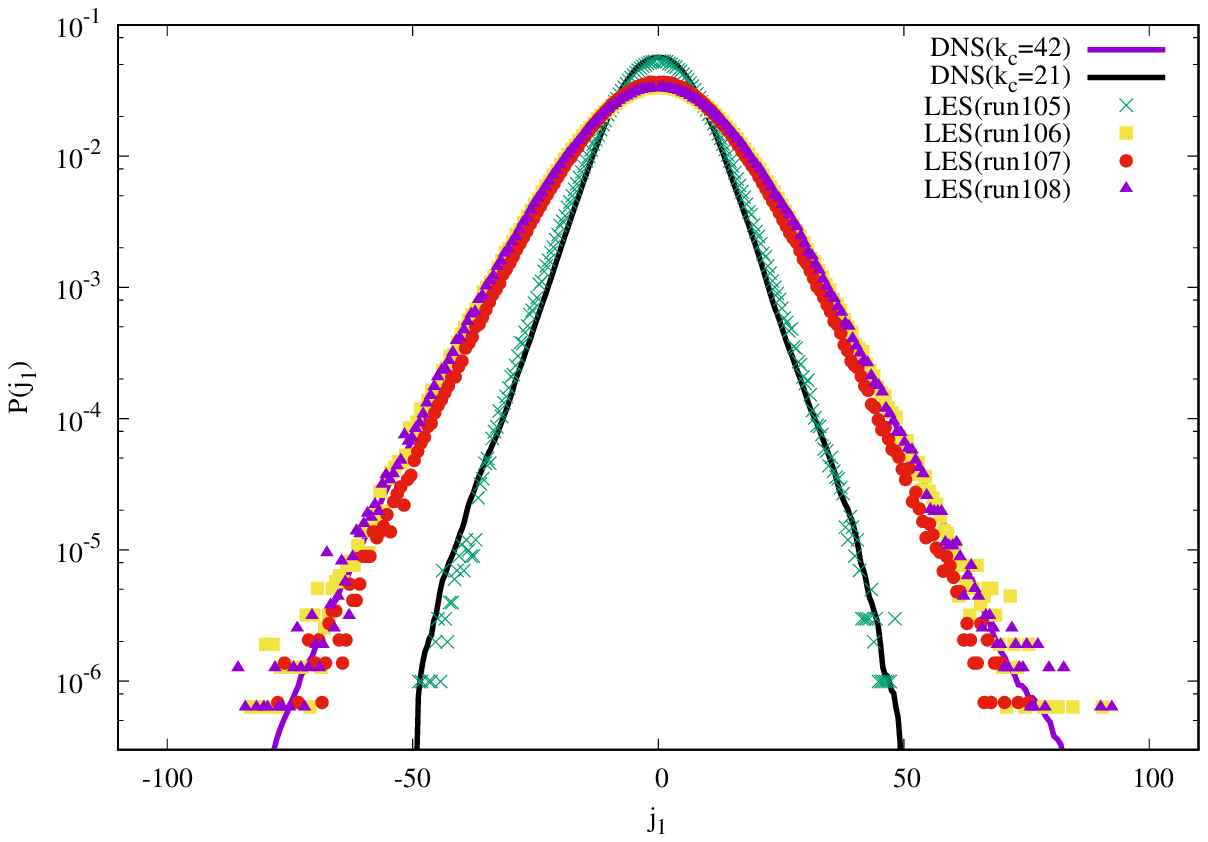}
$\;\;$ \\
$\;\;$ \\
\caption{\label{fig:LESpdf01} A comparison of the PDFs of $1st$-component of (a) the vorticity $\omega_1$ and (b) the current density $j_1$ among runs 100-109 and DNS GS data at $t=1$. }
\end{center}
\end{figure}

\begin{figure}
\begin{center}
\begin{minipage}{0.45\textwidth}
(a)\hspace*{0.9\textwidth}\;\; \\    
  \includegraphics[width=0.99\textwidth]{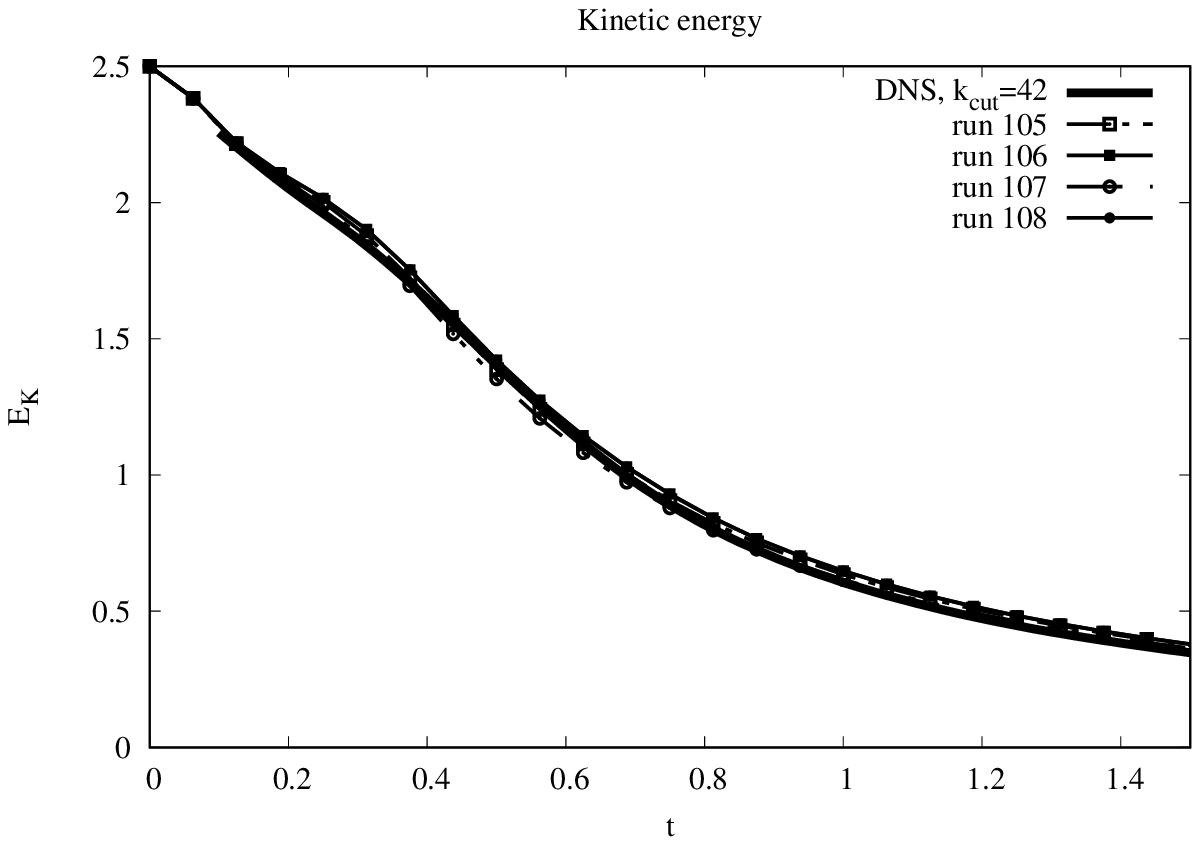}
(b)\hspace*{0.9\textwidth}\;\; \\    
  \includegraphics[width=0.99\textwidth]{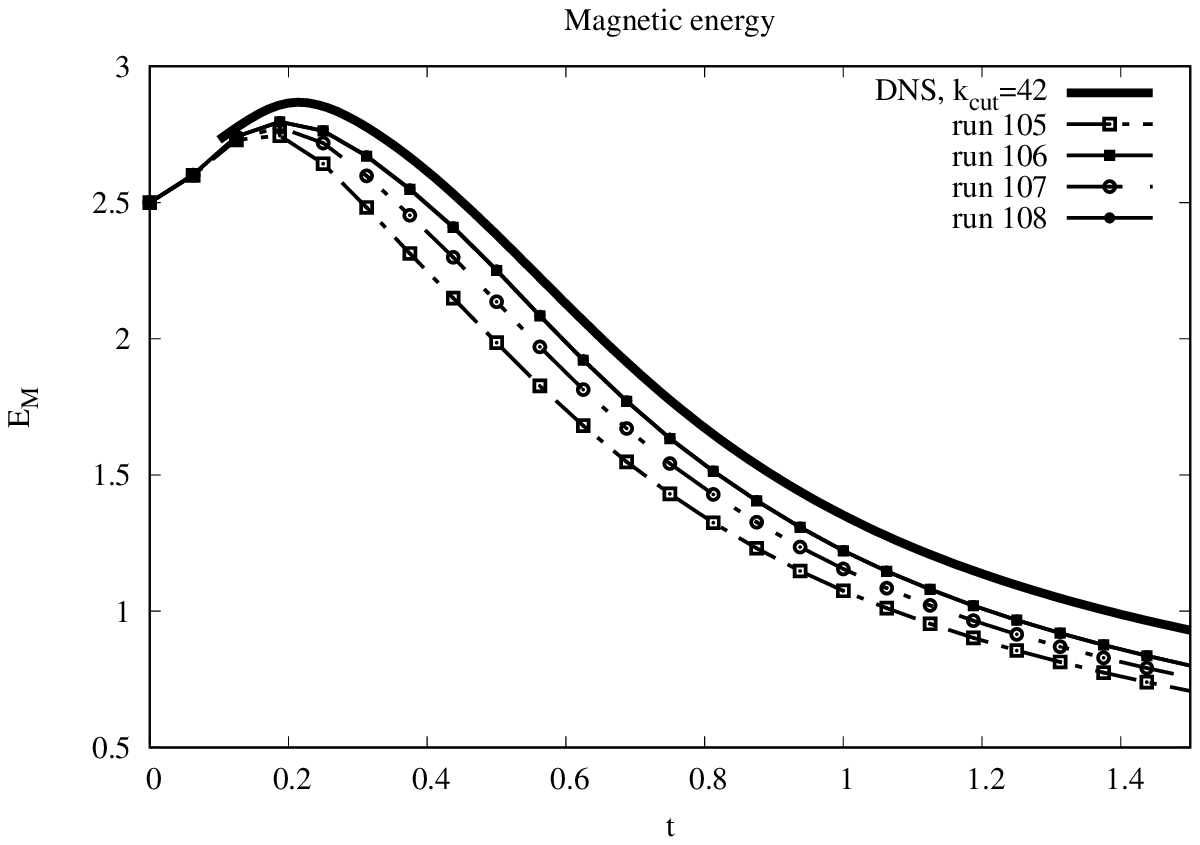}
\end{minipage}
\begin{minipage}{0.45\textwidth}
(c)\hspace*{0.9\textwidth}\;\; \\    
  \includegraphics[width=0.99\textwidth]{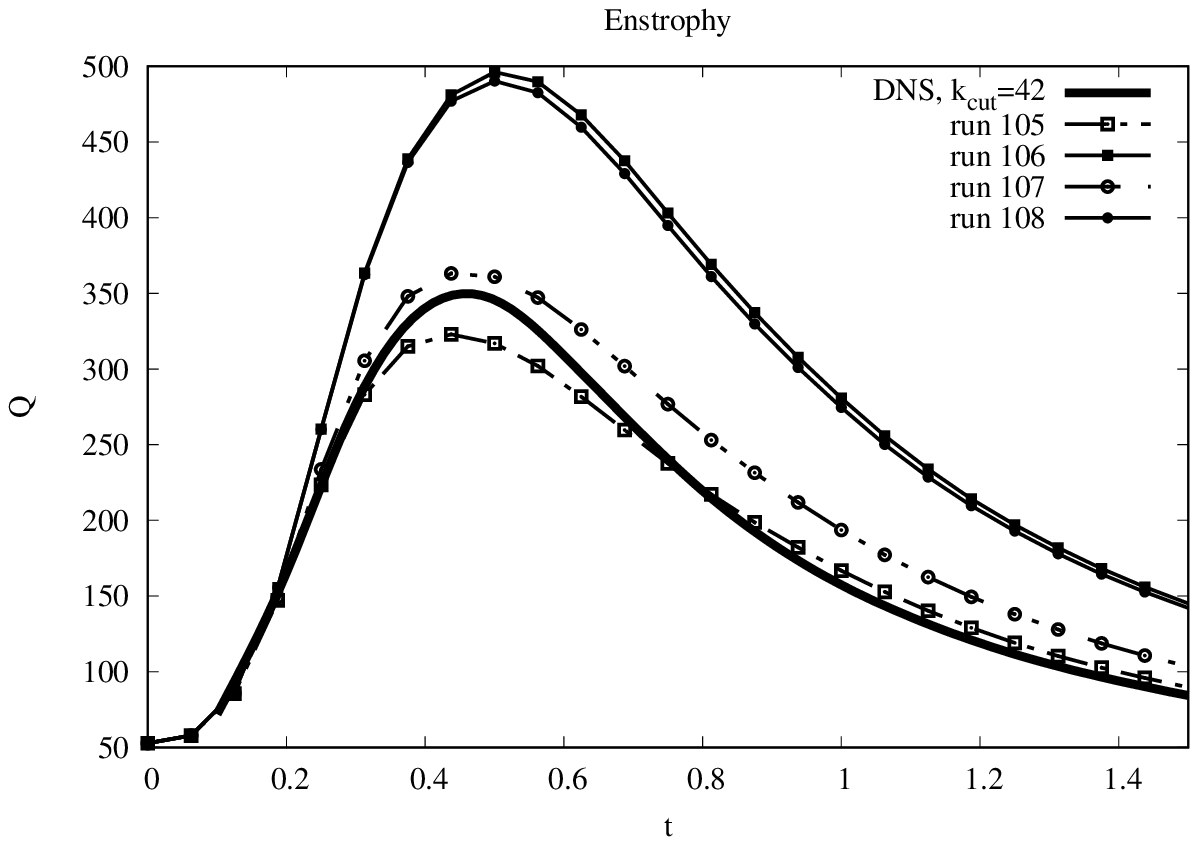}
(d)\hspace*{0.9\textwidth}\;\; \\    
  \includegraphics[width=0.99\textwidth]{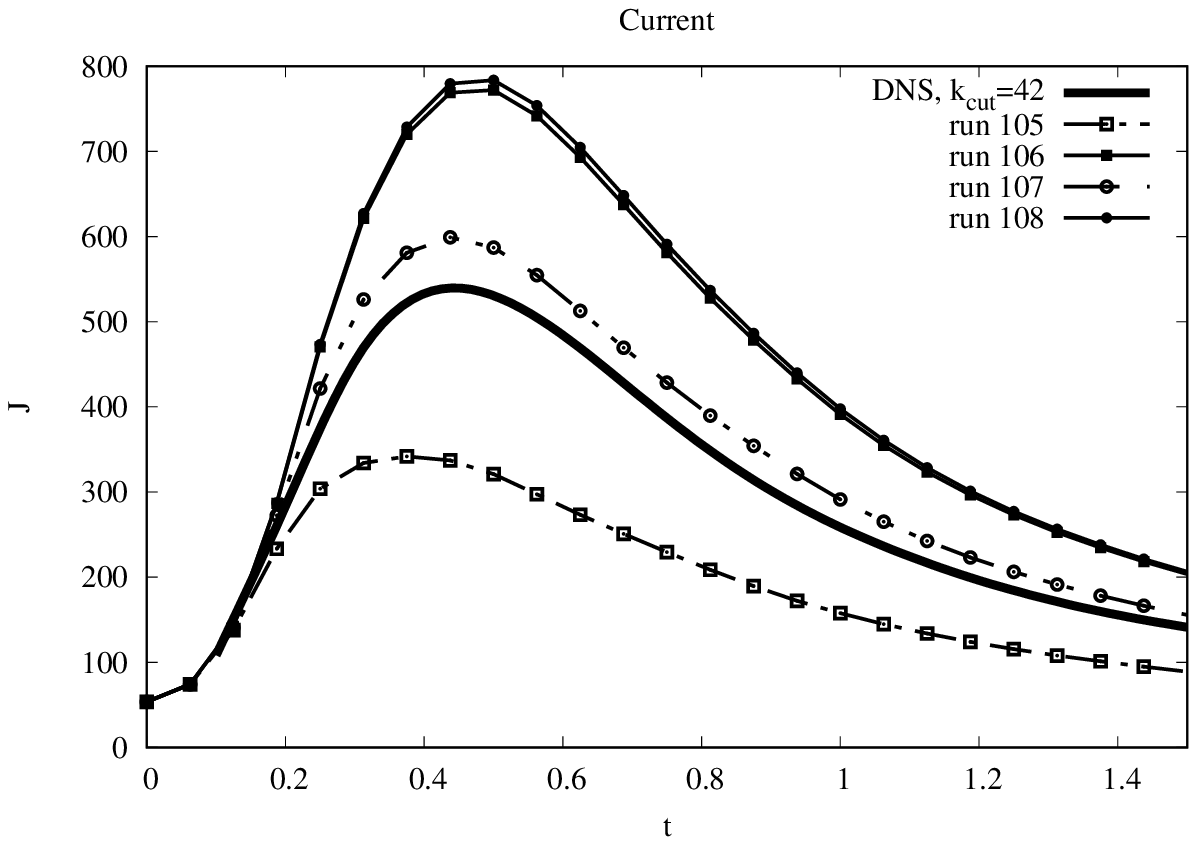}
\end{minipage}
$\;\;$ \\
$\;\;$ \\
\caption{\label{fig:LESstats03} Time evolution of (a) $E_K$, (b) $E_M$, (c) $J$, and (d) $Q$ for LES with the parameter sets 105-108 with the number of grid points $N^3=128^3$ and the Hall parameter $\epsilon_H=0.025$. }
\end{center}
\end{figure}

\begin{figure}
\begin{center}
(a)\hspace*{0.5\textwidth}\;\; \\    
  \includegraphics[width=0.55\textwidth]{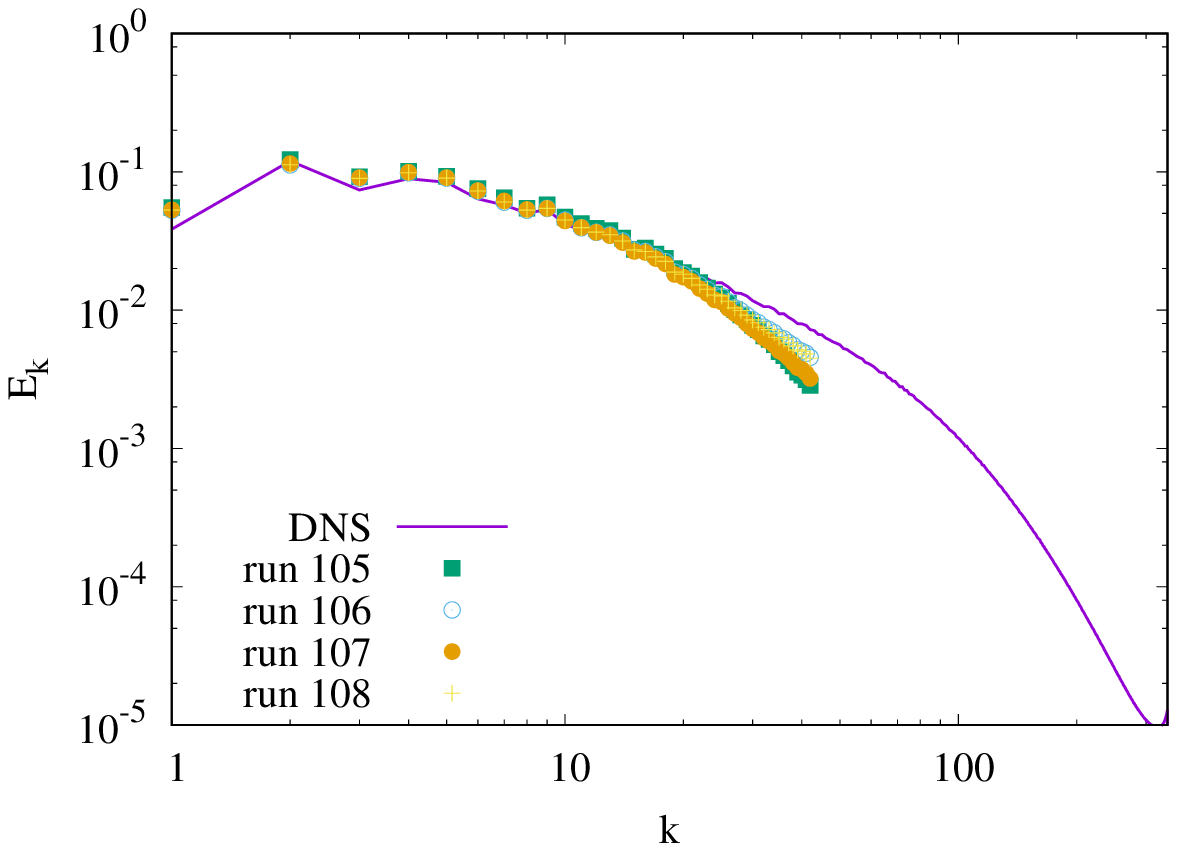}
(b)\hspace*{0.5\textwidth}\;\; \\    
  \includegraphics[width=0.55\textwidth]{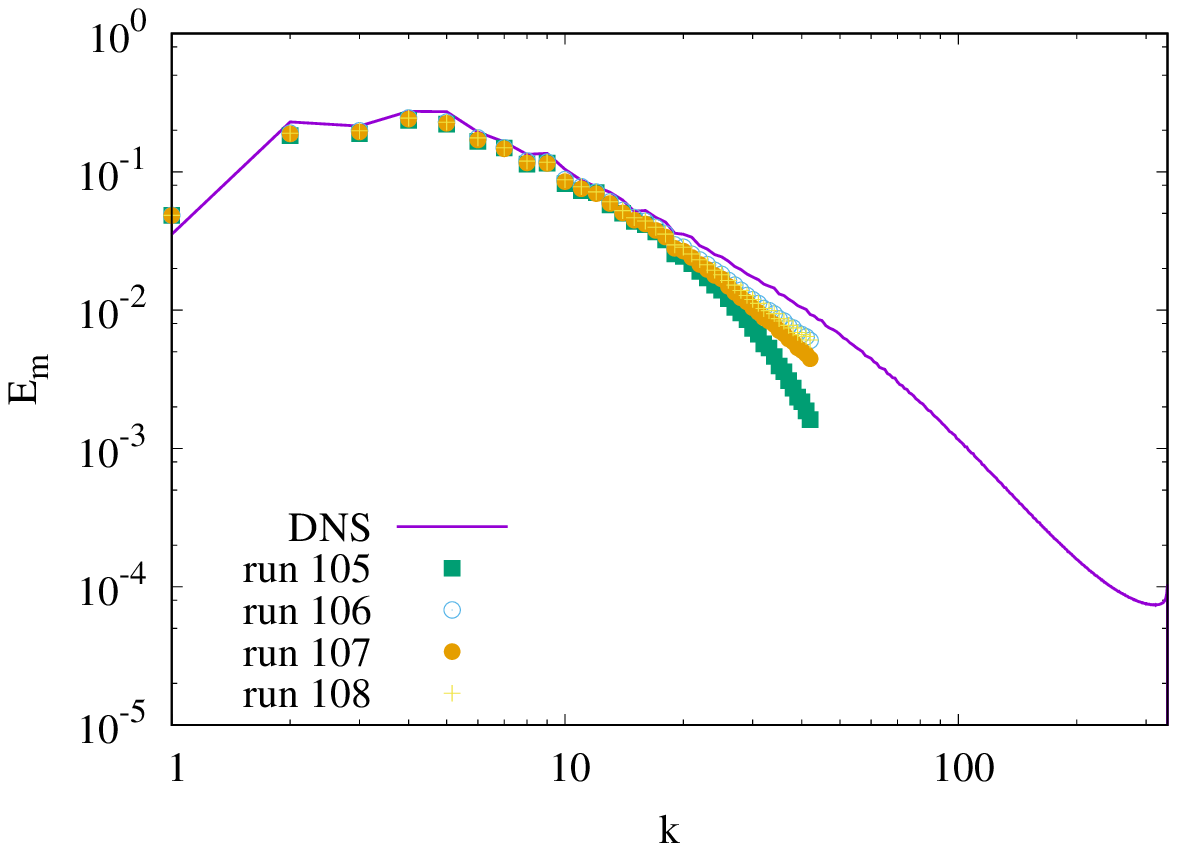}
$\;\;$ \\
$\;\;$ \\
\caption{\label{fig:LESspec03} Energy spectra (a) $E_K(k,t)$ and (b) $E_M(k,t)$ at $t=1$ of runs 105-108 with the number of grid points $N^3=128^3$ and the Hall parameter $\epsilon_H=0.025$. }
\end{center}
\end{figure}

\begin{figure}
\begin{center}
\begin{minipage}{0.45\textwidth}
(a)\hspace*{0.9\textwidth}\;\; \\    
  \includegraphics[width=0.99\textwidth]{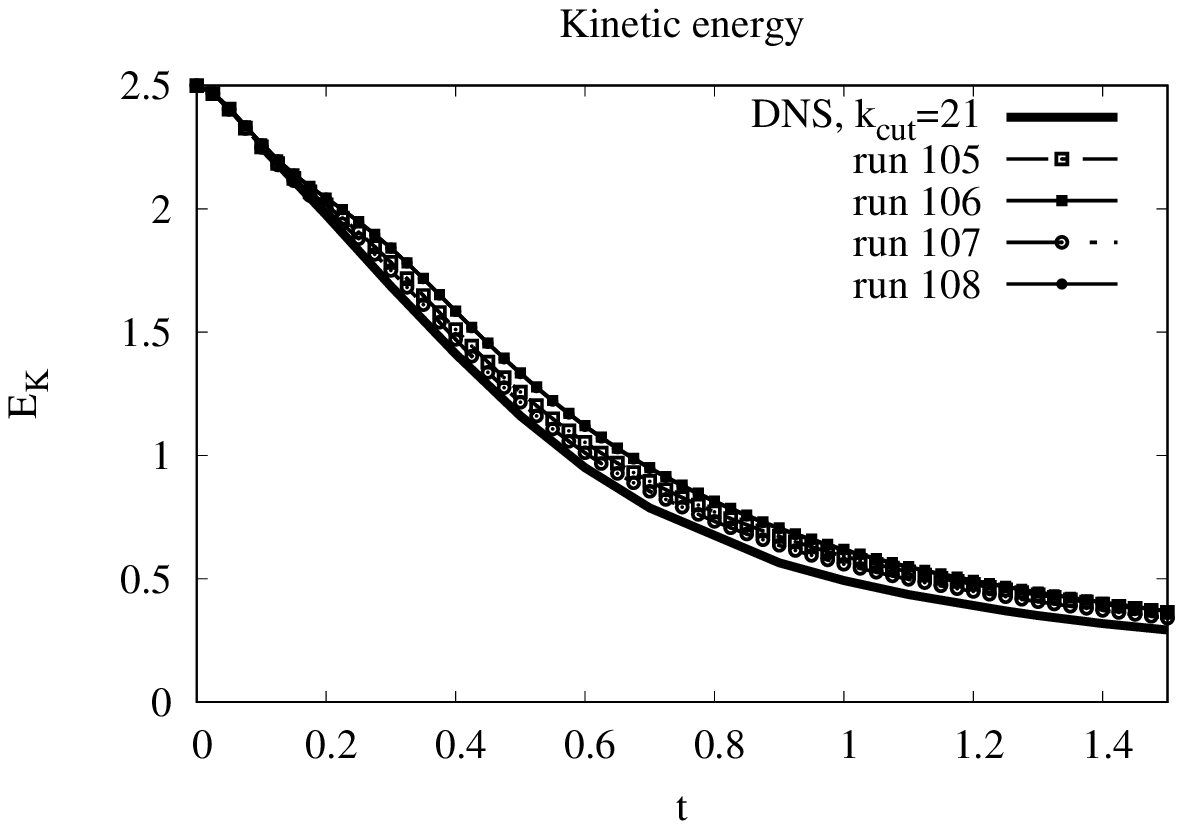}
(b)\hspace*{0.9\textwidth}\;\; \\    
  \includegraphics[width=0.99\textwidth]{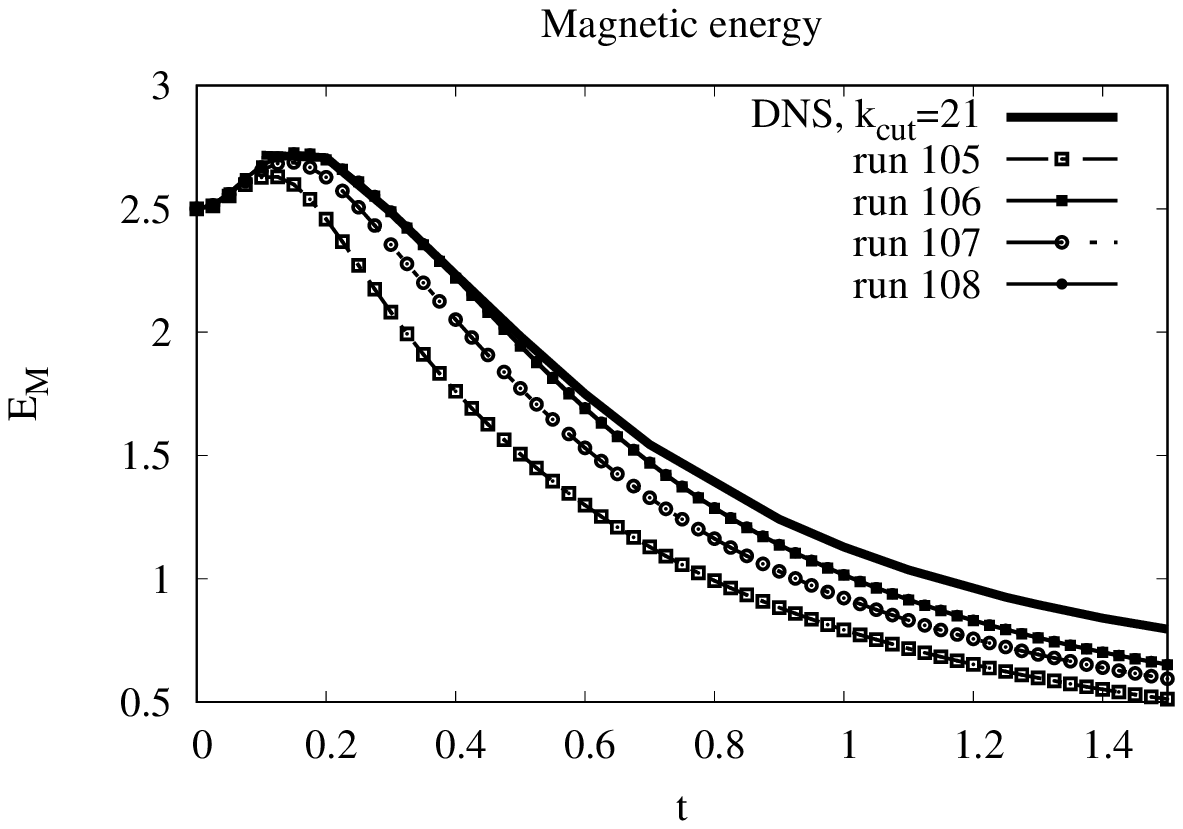}
\end{minipage}
\begin{minipage}{0.45\textwidth}
(c)\hspace*{0.9\textwidth}\;\; \\    
  \includegraphics[width=0.99\textwidth]{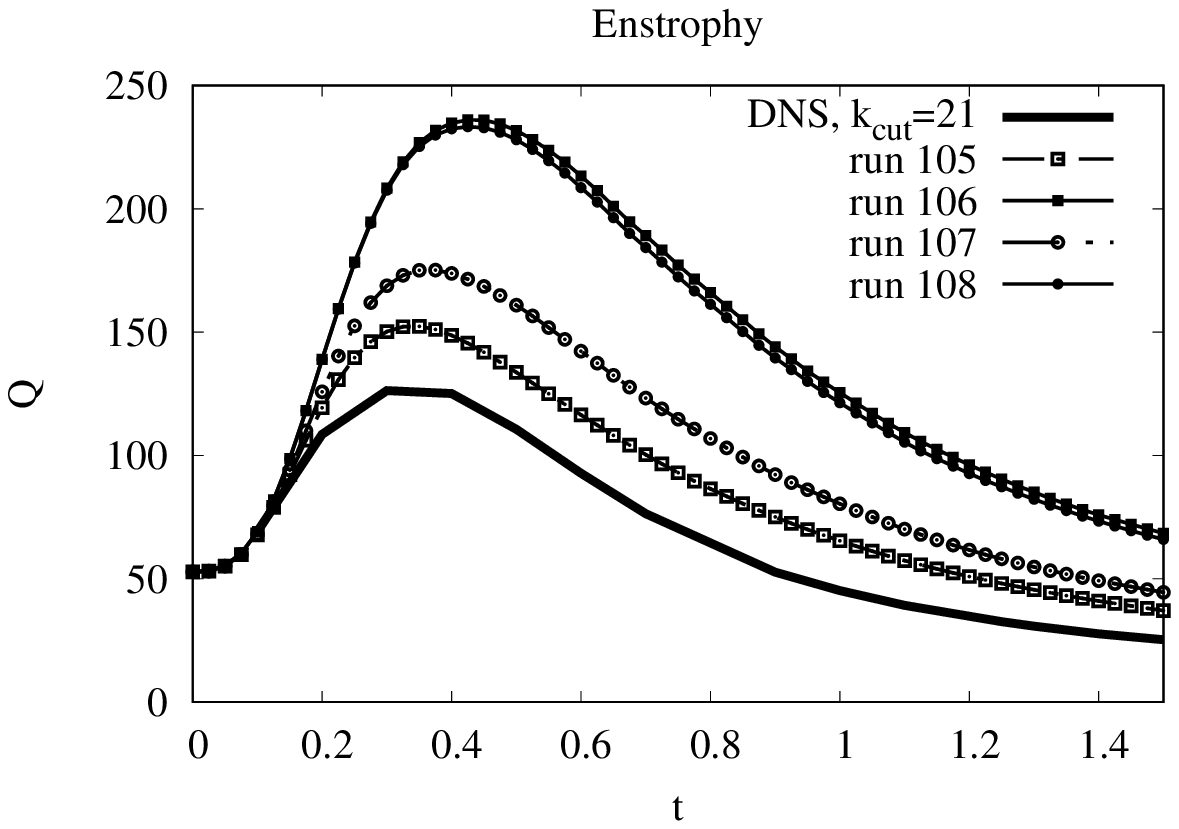}
(d)\hspace*{0.9\textwidth}\;\; \\    
  \includegraphics[width=0.99\textwidth]{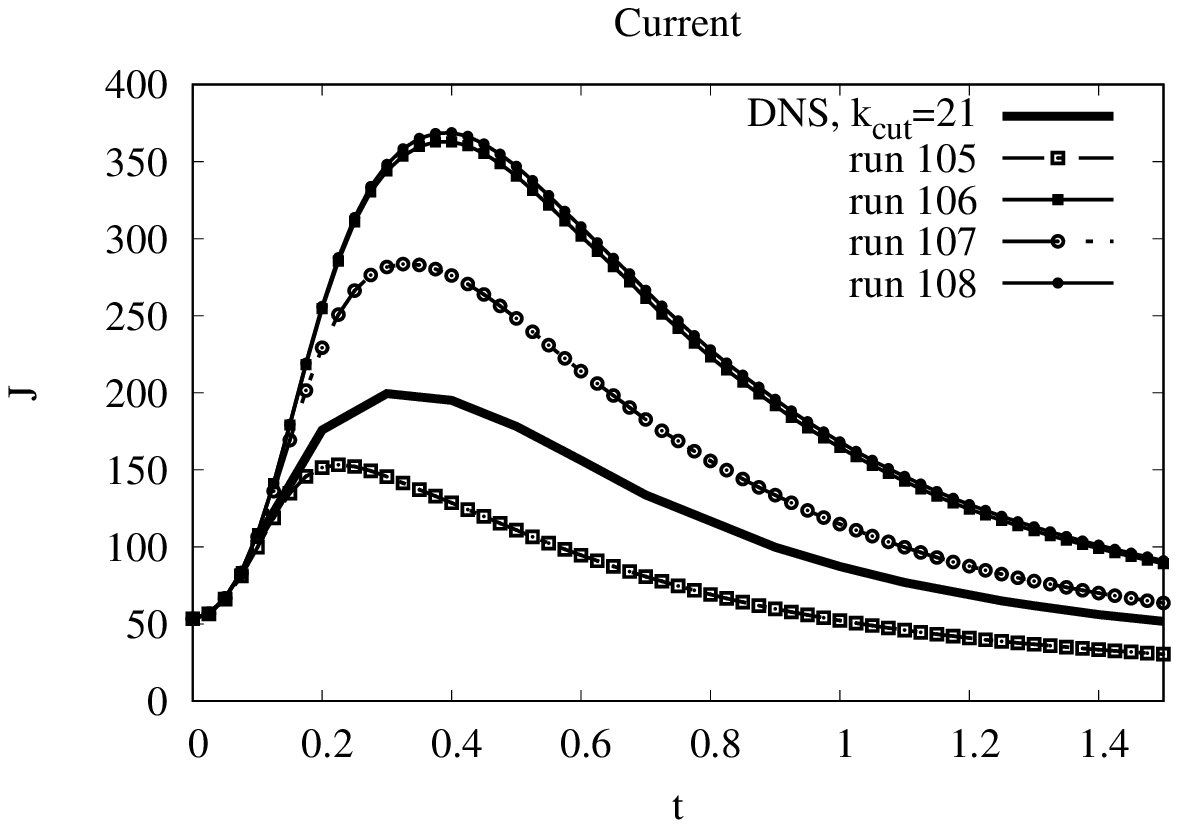}
\end{minipage}
$\;\;$ \\
$\;\;$ \\
\caption{\label{fig:LESstats04} Time evolution of (a) $E_K$, (b) $E_M$, (c) $J$, and (d) $Q$ for LES with the parameter sets 105-108 with the number of grid points $N^3=64^3$ and the Hall parameter $\epsilon_H=0.05$. }
\end{center}
\end{figure}

\begin{figure}
\begin{center}
(a)\hspace*{0.5\textwidth}\;\; \\    
\includegraphics[width=0.55\textwidth]{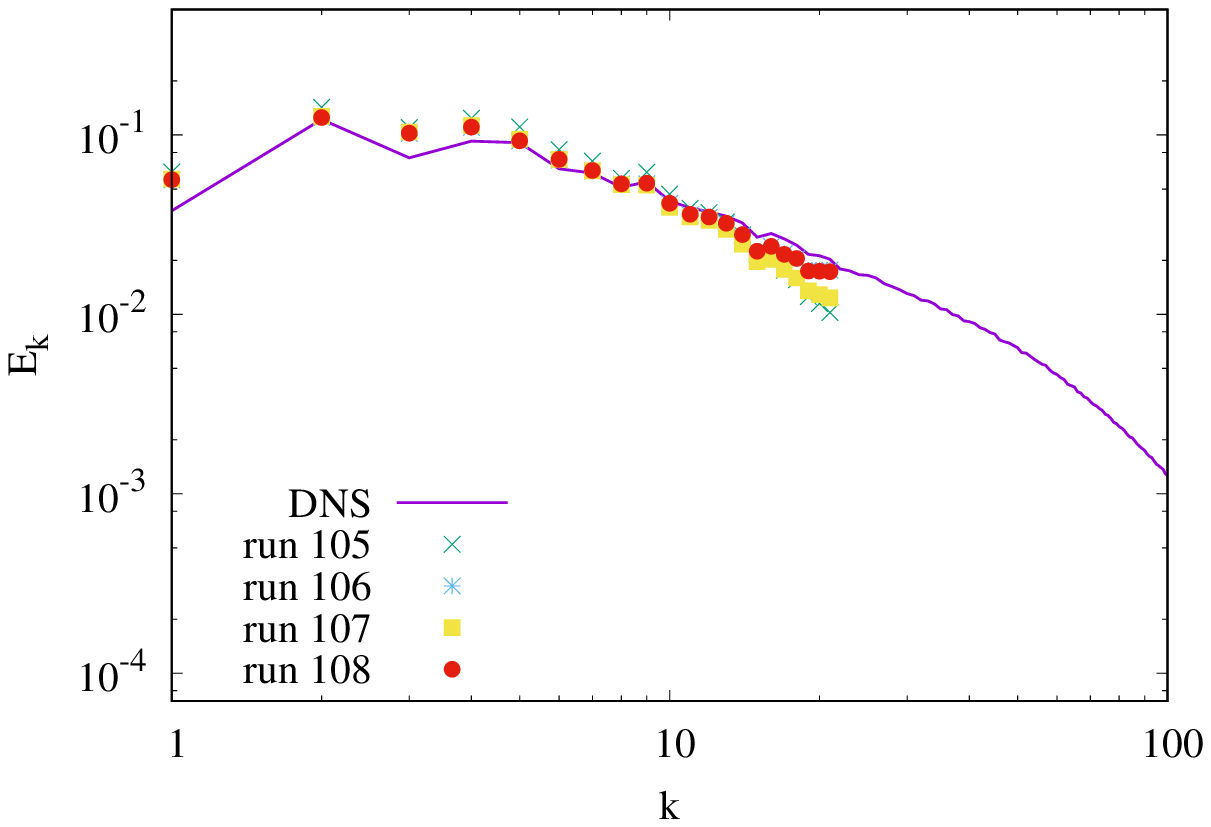}
(b)\hspace*{0.5\textwidth}\;\; \\    
\includegraphics[width=0.55\textwidth]{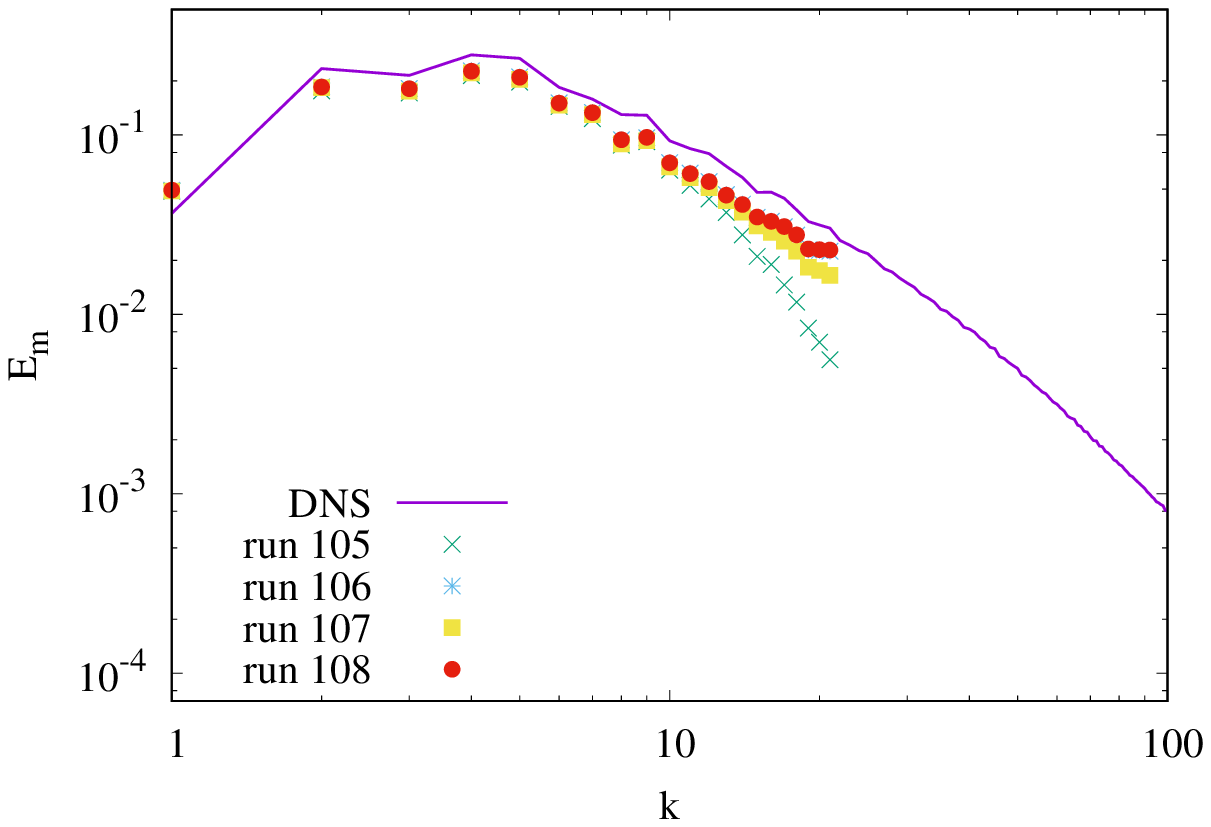}
$\;\;$ \\
$\;\;$ \\ 
\caption{\label{fig:LESspec04} Energy spectra (a) $E_K(k,t)$ and (b) $E_M(k,t)$ at $t=1$ of runs 105-108 with the number of grid points $N^3=64^3$.}
\end{center}
\end{figure}

\begin{description}
\item[Figure \ref{fig:LESstats01}]$\;$ \\ \noindent
Time evolution of (a) $E_K$, (b) $E_M$, (c) $Q$, and (d) $J$ of runs 000-004. 

\item[Figure \ref{fig:LESspec01}] $\;$ \\ \noindent
Energy spectra (a) $E_K(k,t)$ and (b) $E_M(k,t)$ at $t=1$ of runs 000-004.

\item[Figure \ref{fig:LESiso00}] $\;$ \\ \noindent
Isosurfaces of $q$ (with blue-green colors) and $I$ (gray) at $t=1$ in LES of the runs (a) 000 and (b) 003.
  
\item[Figure \ref{fig:LESstats02}] $\;$ \\ \noindent
Time evolution of (a) $E_K$, (b) $E_M$, (c) $J$, and (d) $Q$ of runs 100-109. 

\item[Figure \ref{fig:LESspec02}] $\;$ \\ \noindent
Energy spectra (a) $E_K(k,t)$ and (b) $E_M(k,t)$ at $t=1$ of runs 100-104, and (c) $E_K(k,t)$ and (d) $E_M(k,t)$ at $t=1$ of runs 105-109.

\item[Figure \ref{fig:LESiso01}] $\;$ \\ \noindent
  Isosurfaces of $q$ (with blue-green colors) and $I$ (gray) at $t=1$ in LES of the runs (a) 105, (b) 106, (c) 107, and (d) 108.
  
\item[Figure \ref{fig:LESpdf01}] $\;$ \\ \noindent
  A comparison of the PDFs of $1st$-component of (a) the vorticity $\omega_1$ and (b) the current density $j_1$ among runs 100-109 and DNS GS data at $t=1$.

\item[Figure \ref{fig:LESstats03}] $\;$ \\ \noindent
  Time evolution of (a) $E_K$, (b) $E_M$, (c) $J$, and (d) $Q$ for LES with the parameter sets 105-108 with the number of grid points $N^3=128^3$ and the Hall parameter $\epsilon_H=0.025$.

\item[Figure \ref{fig:LESspec03}] $\;$ \\ \noindent
  Energy spectra (a) $E_K(k,t)$ and (b) $E_M(k,t)$ at $t=1$ of runs 105-109 with the number of grid points $N^3=128^3$ and the Hall parameter $\epsilon_H=0.025$.
  
\item[Figure \ref{fig:LESstats04}] $\;$ \\ \noindent
  Time evolution of (a) $E_K$, (b) $E_M$, (c) $J$, and (d) $Q$ for LES with the parameter sets 105-109 with the number of grid points $N^3=64^3$ and the Hall parameter $\epsilon_H=0.05$.
  
\item[Figure \ref{fig:LESspec04}] $\;$ \\ \noindent
  Energy spectra (a) $E_K(k,t)$ and (b) $E_M(k,t)$ at $t=1$ of runs 105-109 with the number of grid points $N^3=64^3$.

\end{description}

\clearpage

\begin{table}[htb]
\begin{center}
  \caption{Parameters for LES of Hall MHD turbulence.}
%
\begin{tabular}{|c|c|c|c|c|c|c|c|c|c|c|}
\hline
\multirow{2}{*}{Run No.}
         & \multicolumn{2}{|c}{$\Delta=\Delta_0$}  
         & \multicolumn{2}{|c}{$\Delta=1.5\Delta_0$} 
         & \multicolumn{2}{|c|}{$\Delta=2\Delta_0$} 
         & \multicolumn{2}{|c|}{$\Delta=3\Delta_0$} 
         & \multirow{2}{*}{$\sigma_{{}_{H,2}}\;$} 
         & \multirow{2}{*}{$\sigma_{{}_{H,2}}\;$} \\ \cline{2-9}
         & $C_{\nu}$  & $C_{\eta}$   & $C_{\nu}$   & $C_{\eta}$ & $C_{\nu}$   & $C_{\eta}$ & $C_{\nu}$  & $C_{\eta}$ &     &       \\ \hline \hline
   000   & $0.092$    & $0.092$      & $0.053580$  & $0.053580$ & $0.036510$  & $0.036510$ & $0.021263$ & $0.021263$ & $0$ & $0$   \\ \hline
   001   & $0.13$     & $0.13$       & $0.075710$  & $0.075710$ & $0.051591$  & $0.051591$ & $0.030046$ & $0.030046$ & $0$ & $0$   \\ \hline
   002   & $0.046$    & $0.046$      & $0.026790$  & $0.026790$ & $0.018255$  & $0.018255$ & $0.010632$ & $0.010632$ & $0$ & $0$   \\ \hline
   003   & $0.23$     & $0.23$       & $0.133949$  & $0.133949$ & $0.091276$  & $0.091276$ & $0.053158$ & $0.053158$ & $0$ & $0$   \\ \hline
   004   & $0.345$    & $0.345$      & $0.200924$  & $0.200924$ & $0.136913$  & $0.136913$ & $0.079737$ & $0.079737$ & $0$ & $0$   \\ \hline
   100   & $0.092$    & $0.065714$   & $0.053580$  & $0.038271$ & $0.036510$  & $0.026079$ & $0.021263$ & $0.015188$ & $2$ & $0$   \\ \hline
   101   & $0.13$     & $0.092857$   & $0.075710$  & $0.054079$ & $0.051591$  & $0.036850$ & $0.030046$ & $0.021461$ & $2$ & $0$   \\ \hline
   102   & $0.276$    & $0.197143$   & $0.160739$  & $0.114813$ & $0.109531$  & $0.078236$ & $0.063789$ & $0.045564$ & $2$ & $0$   \\ \hline
   103   & $0.23$     & $0.180714$   & $0.133949$  & $0.105246$ & $0.091276$  & $0.071717$ & $0.053158$ & $0.041767$ & $2$ & $0$   \\ \hline
   104   & $0.092$    & $0.065714$   & $0.053580$  & $0.038271$ & $0.036510$  & $0.026079$ & $0.021263$ & $0.015188$ & $2$ & $1$   \\ \hline
   105   & $0.13$     & $0.092857$   & $0.075710$  & $0.054079$ & $0.051591$  & $0.036850$ & $0.030046$ & $0.021461$ & $2$ & $1$   \\ \hline
   106   & $0.092$    & $0.065714$   & $0.053580$  & $0.038271$ & $0.036510$  & $0.026079$ & $0.021263$ & $0.015188$ & $4$ & $0$   \\ \hline
   107   & $0.13$     & $0.092857$   & $0.075710$  & $0.054079$ & $0.051591$  & $0.036850$ & $0.030046$ & $0.021461$ & $4$ & $0$   \\ \hline
   108   & $0.092$    & $0.065714$   & $0.053580$  & $0.038271$ & $0.036510$  & $0.026079$ & $0.021263$ & $0.015188$ & $8$ & $1$   \\ \hline
   109   & $0.23$     & $0.164286$   & $0.133949$  & $0.095678$ & $0.091276$  & $0.065197$ & $0.053158$ & $0.037970$ & $8$ & $0$   \\ \hline
\hline
\end{tabular}
\label{tab:params}
\end{center}
\end{table}

\clearpage

\clearpage

\clearpage

\clearpage

\clearpage

\clearpage

\clearpage

\clearpage

\clearpage

\clearpage

\clearpage

\end{document}